\documentclass[sigconf]{acmart}
\usepackage[utf8]{inputenc}
\usepackage{tikz}
\usepackage{amsmath}
\usepackage{filecontents}
\usepackage[flushleft]{threeparttable}
\usepackage{amsmath,amsfonts}

\usepackage{graphicx}
\usepackage{textcomp}

 \usepackage{url}

\usepackage{lipsum,tabularx}

\usepackage{multicol}
\usepackage{multirow}

\usepackage{colortbl}
\usepackage{booktabs}
\usepackage{setspace}

\hypersetup{
  linkcolor={blue!70!black},
  citecolor={red!70!black},
  urlcolor={blue!70!black}
}
\def\Snospace~{\S{}}

\usepackage[T1]{fontenc}

\usepackage{algorithm}
\usepackage{algorithmicx}
\usepackage[noend]{algpseudocode}
\usepackage{balance}

\usepackage{bm}
\usepackage{fp}
\usepackage{siunitx}
\usepackage{amsthm}

\usepackage[labelfont=bf,font=small,skip=5pt]{caption}
\usepackage{subcaption}

\usepackage{comment}

\usepackage{fancyhdr}
\usepackage{tikz}

\usepackage{xspace}

\newcommand{\boxbeg}{
  \vspace{2px}
  \noindent\begin{tabular}{|l|}\hline
    \begin{minipage}{3.2in}
      \vspace{2px}
      \noindent
      }

      \newcommand{\boxend}{
      \vspace{2px}
    \end{minipage} \\ \hline
  \end{tabular}
  \vspace{-10pt}
}

\usepackage{algpseudocode}

\newcommand{\coo}{\ensuremath{\mathrm{CO_2}}}

\AtBeginDocument{%
  \providecommand\BibTeX{{%
    \normalfont B\kern-0.5em{\scshape i\kern-0.25em b}\kern-0.8em\TeX}}}

\setcopyright{acmcopyright}
\copyrightyear{2018}
\acmYear{2023}

\setcopyright{none}
\renewcommand\footnotetextcopyrightpermission[1]{} 

\begin{document}

\newcommand{\sys}{\mbox{\textsc{DREAM}}\xspace}
\title{\sys: Debugging and Repairing AutoML Pipelines}

\pagestyle{plain}

\author{Xiaoyu Zhang}
\affiliation{%
  \institution{Xi'an Jiaotong University}
  \city{Xi'an}
  \country{China}
}
\email{zxy0927@stu.xjtu.edu.cn}

\author{Juan Zhai}
\affiliation{%
  \institution{University of Massachusetts}
  \country{United States}
}
\email{juanzhai@umass.edu}

\author{Shiqing Ma}
\affiliation{%
  \institution{University of Massachusetts}
  \country{United States}
}
\email{shiqingma@umass.edu}

\author{Chao Shen}
\affiliation{%
  \institution{Xi'an Jiaotong University}
  \city{Xi'an}
  \country{China}
}
\email{chaoshen@xjtu.edu.cn}

\begin{abstract}
    Deep Learning models have become an integrated component of modern software systems.
    In response to the challenge of model design, researchers proposed Automated Machine Learning (AutoML) systems, which automatically search for model architecture and hyperparameters for a given task.
    Like other software systems, existing AutoML systems suffer from bugs.
    We identify two common and severe bugs in AutoML, \textit{performance bug} (i.e., searching for the desired model takes an unreasonably long time) and \textit{ineffective search bug} (i.e., AutoML systems are not able to find an accurate enough model).
    After analyzing the workflow of AutoML, we observe that existing AutoML systems overlook potential opportunities in search space, search method, and search feedback, which results in performance and ineffective search bugs.
    Based on our analysis, we design and implement \sys, an automatic debugging and repairing system for AutoML systems.
    It monitors the process of AutoML to collect detailed feedback and automatically repairs bugs by expanding search space and leveraging a feedback-driven search strategy.
    Our evaluation results show that \sys can effectively and efficiently repair AutoML bugs.
\end{abstract}

  

\maketitle

\section{Introduction}\label{s:introduction}
In the Software 2.0 era, Machine Learning (ML) techniques that power the intelligent components of a software system are playing an increasingly significant role in software engineering.
The development of Deep Learning (DL) brings software intelligence broader prospects with its recent advances, and DL models are increasingly becoming an integral part of the software systems.
The global edge AI software market is predicted to grow from \$590 million in 2020 to \$1,835 million by 2026~\cite{marketsandmarkets}.
The COVID-19 pandemic also accelerates the application of DL techniques in various industries.
Microsoft Azure Health Bot service helps hospitals classify patients and answer questions about symptoms~\cite{microsoftcovid19}.
Google and Harvard Global Health Institute release the improved COVID-19 Public Forecasts based on AI techniques to provide a projection of COVID-19 cases~\cite{googlecovid19}.
As a growing trend, DL techniques are studied in a wide range of industries, involving domain experts to use DL to solve domain or even task-specific problems.
Unfortunately, most domain experts have limited or no knowledge of DL.
This raises a great challenge for the software engineering community.

In response to the challenge, researchers proposed the concept of Automated Machine Learning (AutoML), which aims to design and build ML models without domain knowledge but only the provided task.
In other words, it takes the user-provided tasks as inputs and automatically builds an ML model that suits the task.
At present, many AI companies and institutions have built and publicly shared AutoML systems (e.g., AutoKeras~\cite{AutoKeras}, NNI~\cite{MicrosoftNNI}, and Cloud AutoML~\cite{cloudAutoML}) to assist users with little or no DL knowledge to build high-quality models.
The workflow of the AutoML pipeline is as follows.
First, AutoML engines preprocess the data and extract features.
Then, they define the search space and design the search strategy to look for possible models and hyperparameters.
These generated models are then trained and evaluated by the engine.
If the validation accuracy is high enough, the process terminates.
Otherwise, it continues to search for the next possible model architecture and hyperparameters.
This process (i.e., generating and training models, evaluating models) iteratively continues until a model meets the predefined condition, typically a threshold for validation accuracy.
This AutoML design has been shown to be effective and efficient in many fields.
For example, the AutoML system powers healthcare and drastically reduces the processing time of medical diagnosis~\cite{AutoML_healthcare}.
It also supports data-driven publishing and reduces the time overhead of content classification from years to months~\cite{AutoML_Digital}.

Like other software systems, AutoML suffers from bugs.
The performance bug (searching for a long time without finishing) and the ineffective search bug (searched architecture and hyperparameters are useless) are the most common bugs in AutoML, which lead to low effectiveness and poor performance~\cite{xie2017genetic,zoph2016neural,he2021automl}.
To understand the root cause of these bugs, we manually debug AutoKeras~\cite{AutoKeras}, one of the most popular AutoML engines.
We find that existing AutoML engines ignore several potential optimization opportunities, which lead to performance and ineffective search bugs.
Firstly, the search space of current pipelines misses some valuable optimization options in training, making the search difficult to find optimal models.
That is, values of many searchable parameters are fixed in existing engines.
We find that changing them can benefit the pipeline.
Secondly, existing search strategies can still be improved.
They often have a low probability of predicting the optimal architecture/hyperparameter change, and require a large number of search trials to obtain a model with desired performance.
This directly leads to the waste of time and resources in search.
Last but not least, existing AutoML engines leverage the validation accuracy as the only feedback to guide the search, which is insufficient to evaluate the quality of previous searches.
This also makes it impossible to identify potential problems in existing models.
As a result, the existing AutoML pipeline requires hundreds of trials to search for a model even on simple datasets, and the obtained model can have far lower accuracy than expected.

In this paper, we propose and implement \sys to repair the performance and ineffective search bugs in the AutoML pipeline.
\sys monitors the model training and evaluation in AutoML pipelines to spot potential bugs.
If a bug is found, \sys fixes them through mechanisms supplemented on the pipeline, namely the search space expansion and the feedback-driven search.
That is, \sys expands the search space and adds effective training options for the model search, such as weight initializers and learning rate scheduling strategy, whose positive effects on model performance have been demonstrated in existing research~\cite{klambauer2017self,he2015delving,loshchilov2017decoupled,qian1999momentum}.
\sys also leverages a novel feedback-driven search strategy to effectively search for model architectures and hyperparameters.
Unlike the simple feedback in the existing pipeline that ignores training traces and the model itself, \sys records many types of feedback (e.g., model loss, gradients) from the model training and evaluation.
Our search will leverage such observations to determine optimal actions, i.e., how to modify the model architectures and hyperparameters (in the expanded search space).
By doing so, \sys effectively fixes performance and ineffective search bugs.

In summary, our contributions can be categorized as follows:
\begin{itemize}
    \item We debug the existing AutoML pipeline and identify the root causes of the performance and ineffective search bugs of the pipeline.
    \item We propose and design the three mechanisms, i.e., search space expansion, feedback-driven search, and feedback monitoring, to detect and repair AutoML performance and ineffective search bugs.
	\item We develop a prototype \sys based on the proposed ideas and evaluate it on four public datasets.
    The results show that, on average, models repaired by \sys achieve 83.21\% accuracy, while the best result of using existing methods is 54.66\%.
	\item Our implementation and data are publicly available at~\cite{ourrepo}.
\end{itemize}

\smallskip
\noindent
{\bf Threat to Validity.}
\sys is currently evaluated on four datasets and eighteen searches, which may be limited.
This evaluation is a months-long effort on powerful hardware.
Although our experiments show that the mechanisms in \sys can efficiently and effectively repair AutoML bugs, they may not hold when the number of the datasets and searches are significantly larger.
To mitigate these threats and follow the Open Science Policy, all the search results (e.g., model architectures and hyperparameters) and training configurations, implementations including dependencies, and evaluation data (e.g., search logs) are publicly available at~\cite{ourrepo} for reproduction.
All the necessary information that can be used to reproduce our experiments and the code of the prototype \sys will be released in our repository.


\section{Background \& Related Work}\label{sec:bg}

\subsection{Automated Machine Learning}\label{sec:automl}
\begin{figure}
	\centering
	\includegraphics[width=\linewidth]{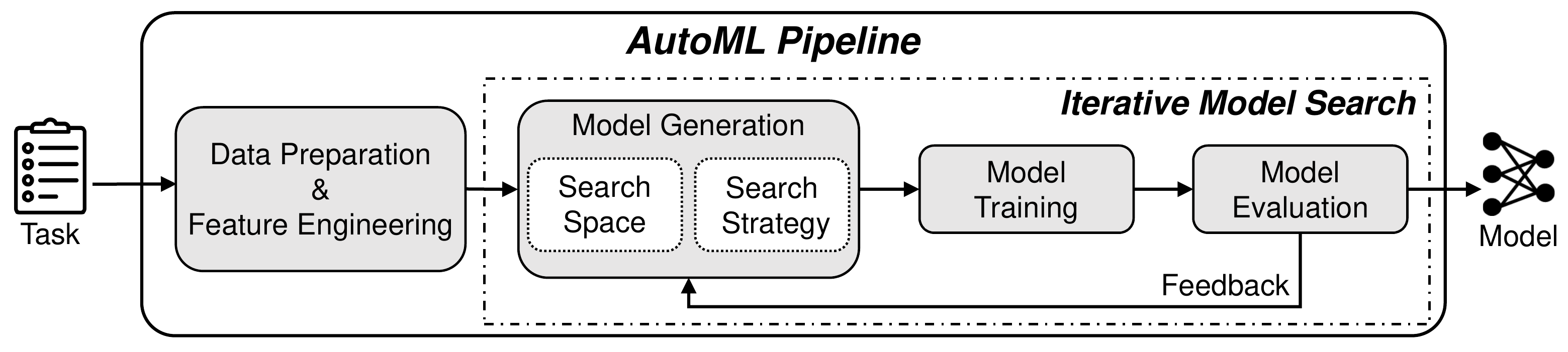}
	\caption{The Basic Steps of the AutoML Pipeline}
	\label{fig:automloverview}
	\vspace{-8pt}
\end{figure}

Designing machine learning (ML) models can be challenging.
In response, researchers propose Automated Machine Learning (AutoML) to build DL models for given datasets without human intervention.
In AutoML, a \textit{search} refers to one attempt to find models for the given task, and each \textit{trial} in a search finds a model to train and evaluate~\cite{he2021automl}.
AutoML engines evaluate every trained model with a \textit{score}, which in most cases is the accuracy of the model.
Moreover, we use \textit{GPU hours} and \textit{GPU days} to measure search efficiency, which are similar concepts to \textit{person-month}.
The AutoML pipeline contains a series of steps~\cite{he2021automl} (shown in \autoref{fig:automloverview}): data preparation and feature engineering, model generation, model training, and model evaluation.
A trial consists of the latter three steps. 
The whole process is iterative until the maximum number of trials is reached or a model with the desired accuracy is found.

\textit{Data preparation and feature engineering} prepare training data and extract features (if necessary) for searches and model building.
The former collects data from users, cleans and normalizes it, and performs data augmentations if needed~\cite{yang2018recognition, roh2019survey, jesmeen2018survey, xie2017data, ho2019population}.
The latter conducts feature engineering, e.g., correlation coefficient, SVM-REF, and principal component analysis (PCA)~\cite{guyon2006introduction,khalid2014survey,huang2014svm}.

\textit{Model generation} undertakes the main task of searching and generating models in the AutoML pipeline.
There are two important components in an AutoML system design, \textit{search space} and \textit{search strategy}.
The \textit{search space} describes all possible model architectures, hyperparameters, and other configurable and searchable optimization options.
AutoML engines then leverage different heuristics or statistical methods, namely \textit{search strategies}, to search for models.
There are two tasks in this step: neural architecture search (NAS) and hyperparameter optimization (HPO).
NAS searches for suitable network architectures for given tasks, and HPO optimizes all configurable parameters~\cite{shin2020hippo} such as learning rates.
Existing search strategies are typically gradient- or surrogate model-based methods~\cite{zela2018towards,dikov2019bayesian,white2019bananas,jin2019auto,luo2018neural,cai2018proxylessnas,dong2019searching,he2020milenas,pedregosa2016hyperparameter,chandra2019gradient}.
They keep all search history and conduct gradient descent or train surrogate models (e.g., Gaussian processes) to optimize towards the desired goal~\cite{liu2018darts,kandasamy2018neural,negrinho2019towards}.
For example, the Bayesian search strategy in AutoKeras~\cite{jin2019auto} builds a surrogate Gaussian process model.
It fits the Gaussian process model with all previous training results to search for the next neural architecture.
The Greedy method uses greedy strategies to start new trials from the current best result~\cite{AutoKeras}, and Hyperband~\cite{li2017hyperband} optimizes this method by adding dynamic resources allocations and early-stopping mechanisms to these trials.

\textit{Model Training and model evaluation} start after completing the model generation.
The pipeline firstly trains the generated model.
Subsequently, the model evaluation component evaluates the models and the corresponding training results and provides feedback for future trials.
In most implementations, the evaluation criterion is model accuracy.
Modern AutoML engines~\cite{AutoKeras,domhan2015speeding,klein2017fast,klein2016learning} also leverage early-stopping mechanisms to improve search efficiency.

\subsection{Model Debugging and Repairing}\label{sec:repair}

Model debugging and repairing are important software engineering tasks~\cite{sears:acl18, zhang2018training, jiang2004editing, taotrader, DBLP:journals/corr/abs-1909-03824,fahmy2022hudd,LAMP}.
Ribeiro et al.~\cite{sears:acl18} produce adversarial examples as training data to debug natural language processing (NLP) models. 
Others~\cite{zhang2018training, jiang2004editing} clean up training data that are wrongly labeled to debug RNN models.
Ma et al.~\cite{ma2018mode} propose differential analysis on inputs to fix model overfitting and underfitting problems.
Zhang et al.~\cite{zhang2021autotrainer} propose a DNN debugging tool that can automate debugging and repairing of 5 kinds of training problems in the models.
Moreover, DeepLocalize~\cite{wardat2021deeplocalize} debugs and identifies the defect of DNN models. 
It analyzes the historical trends of neuron activation values to identify and localize the faults in the model.
HUDD~\cite{fahmy2022hudd} automatically clusters error-inducing images and captures the relevance of these images and the neurons in a DNN model to conduct a safety analysis for the DNN errors.

A great number of testing methods have been proposed to test ML models, such as fuzzing~\cite{odena2019tensorfuzz,guo2018dlfuzz,xie2019deephunter,zhou2019metamorphic,gao2020fuzz}, symbolic execution~\cite{gopinath2019symbolic,usman2021neurospf}, runtime validation~\cite{li2021testing,xiao2021self}, fairness testing~\cite{zheng2021neuronfair,segal2021fairness}, etc.
DeepXplore~\cite{pei2017deepxplore} introduces the neuron coverage metric to measure the percentage of activated neurons or a given test suite and DNN models. 
Furthermore, it generates new test inputs to maximize the metric. 
Others~\cite{ma2018deepgauge, xie2019deephunter, zhang2018deeproad, zhou2020deepbillboard, gerasimou2020importance} extend the coverage concept to different scenarios.
Model testing are useful for other domains such as image classification~\cite{sun2018concolic,tian2020testing}, automatic speech recognition~\cite{du2019deepstellar,asyrofi2021can}, text classification~\cite{udeshi2018automated,huang2021coverage}, and machine translation~\cite{he2020structure,sun2020automatic}.
Huang et al.~\cite{huang2021coverage} utilize the internal behavior of RNNs and develop a coverage criterion to guide the test for the defects in RNN models.
Rossolini et al.~\cite{rossolini2021increasing} present four coverage analysis methods to conduct testing on the adversarial examples and out-of-distribution inputs. 
Yan et al.~\cite{DBLP:conf/sigsoft/YanTLZMX020} measure correlations between coverage and model quality (e.g., model robustness) and empirically show that existing criteria can not reflect model quality.
Yang et al.~\cite{yang2022revisiting} conduct deep analysis in neuron coverage metrics and confirm the conclusion from Yan et al.~\cite{DBLP:conf/sigsoft/YanTLZMX020} that coverage-driven methods are less effective than gradient-based methods.

\begin{figure}
	\centering
    \footnotesize
	\includegraphics[width=0.9\columnwidth]{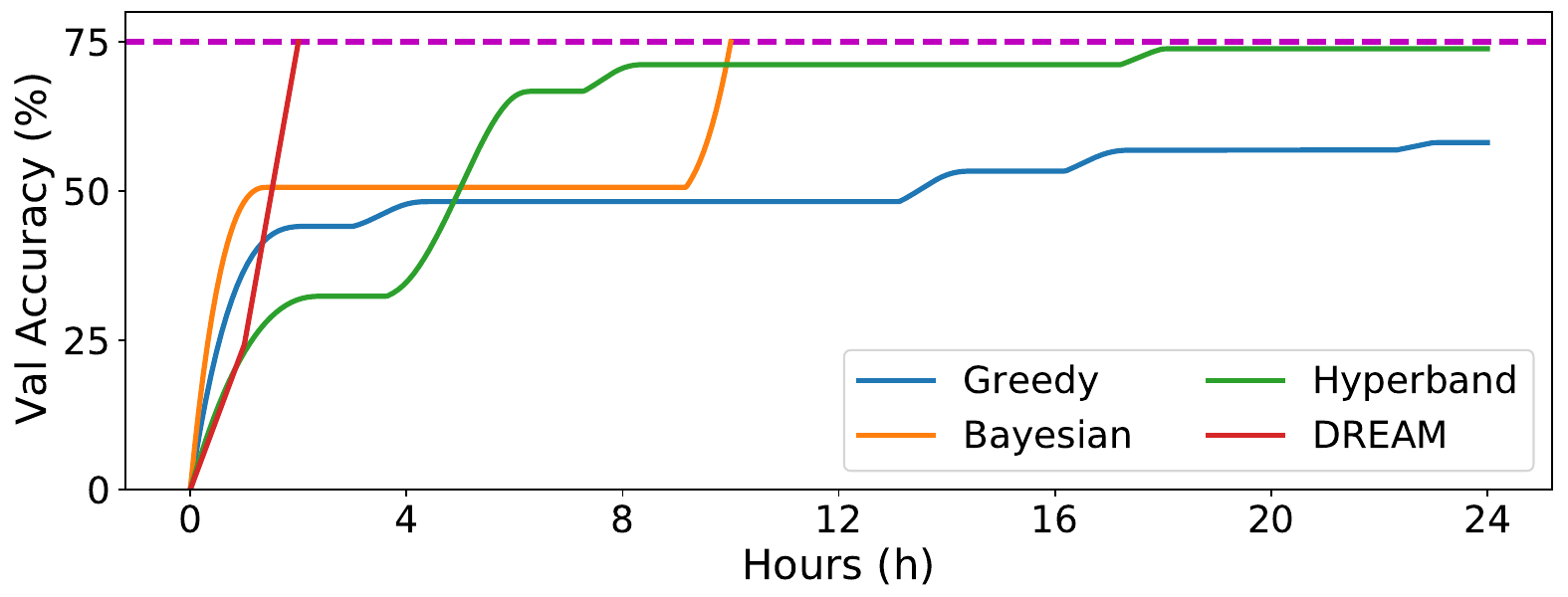}
	\caption{Performance bug in AutoML When Search for Target Score}\label{fig:moti_efficiency}
	\vspace{-8pt}
\end{figure}

\section{Motivation}\label{sec:moti}

Like other systems, AutoML has bugs.
Besides common software system bugs, there are two types of popular issues of existing AutoML engines: \textit{performance bug} and \textit{ineffective search bug}.
The former, \textit{performance bug} indicates that an AutoML engine can take too much time to reach the desired goal, which leads to high carbon footprints, waste of energy, and inefficient use of resources.
Due to the high cost of modern AI model training, this problem is more severe in AutoML than in other systems.
The latter, \textit{ineffective search bug} refers to the problem that AutoML engines are not able to find optimal model architectures or hyperparameters, which is a typical example of functional errors.
Unlike traditional logic-based programs where testing methods can identify functional errors by formally describing the logic of the given program, the results of AutoML have no ground truth, and determining if the result is optimal or not is challenging.
Thus, fixing the problem or repairing the system is inherently challenging.
In this paper, we focus on these two types of AutoML bugs.

\subsection{Motivating Examples}\label{sec:motivatingexample}
Here, we use AutoKeras~\cite{AutoKeras}, one of the most popular AutoML engines with 8.3k stars in GitHub, to illustrate AutoML bugs.
Specifically, we use AutoKeras to train on the CIFAR-100 dataset with three strategies it provides, i.e., Greedy, Bayesian, and Hyperband.

\smallskip
\noindent\textit{Case I\@: Performance Bug.}
In this experiment, we set the goal of validation accuracy to 75\%.
Each strategy starts from the same seed and runs for no less than 24 hours if it has not achieved the goal.
We use default values for all other configurable parameters.
The Bayesian search strategy uses 9.5 hours and meets the goal.
Hyperband and Greedy methods do not reach the target.
After 24 hours and 242 trials, Hyperband obtains 71.2\% accuracy; and Greedy reaches 70.8\% after three days.
As a comparison, our method \sys\ achieves the target accuracy within 95 minutes, which is one-sixth the time used by the Bayesian strategy.
Compared with the best of AutoKeras, \sys\ saves 7.92 GPU hours and 2.80 kWh electricity and emits 1.21 KG less \coo\ and carbon footprints~\cite{ML_CO2_IMPACT,lacoste2019quantifying}. 


\begin{figure}
	\centering
    \footnotesize
	\includegraphics[width=0.9\columnwidth]{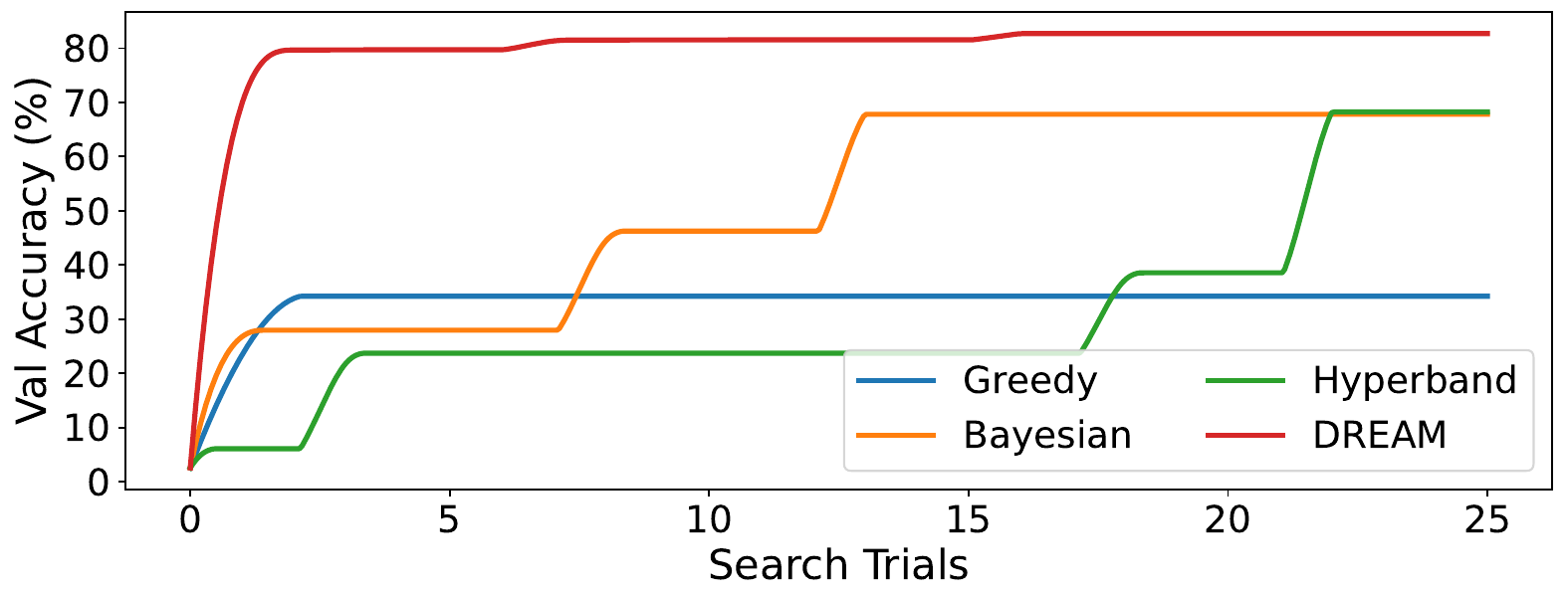}
	\caption{Ineffective Search Trials of the AutoKeras Pipeline} \label{fig:moti_effective}
	\vspace{-8pt}
\end{figure}

\smallskip
\noindent\textit{Case II\@: Ineffective Search Bug.}
In this experiment, we fix the number of trials to be 25 and use default settings for other parameters.
\autoref{fig:moti_effective} shows results of using different search strategies.
Hyperband improves the accuracy in 4/25 trials and achieves 68.21\% accuracy.
Each effective search trial boosts the accuracy by 17\% on average.
Bayesian and Greedy both have three fruitful trials and respectively get 67.79\% and 34.23\% accuracy.
This data demonstrates that most of the trials in existing AutoML pipelines are ineffective, and the accuracy boost achieved by each successful trial is not significant.
As a comparison, our method \sys\ gains 82.66\% accuracy with five productive trials.
Notably, its accuracy keeps increasing even after getting to 81\% (higher than any of AutoKeras methods).

\subsection{Analysis of AutoML}\label{sec:analysis}

To understand the root causes of bugs in AutoML, we compare AutoML results and state-of-the-art manually crafted models, log detailed training information and analyze them.
Through this process, we make three key observations.

\smallskip
\noindent
\textit{Observation I\@: Existing AutoML engines overlook valuable search spaces, which makes them difficult to find optimal models.}
The search space has significant impacts on the performance of the AutoML pipeline~\cite{xie2019exploring}.
Existing AutoML engines have implemented common model architectures and data augmentation techniques, which have shown high effectiveness in practice, to guide the search engine to find optimal models.
Such designs improved the performance of AutoML by prioritizing expert knowledge chosen search space.
Unfortunately, existing engines overlook a set of other possible optimizations, i.e., training configurations such as weight initializers, optimizers, learning rate scheduling strategies which also have remarkable impacts on the output of AutoML~\cite{park2020novel,loshchilov2017decoupled,ge2019step,chollet2021deep}.
In many trials, we find that changing one of the training configurations will lead to a significant accuracy boost.
Thus, expanding the search space can benefit AutoML engines.

\smallskip
\noindent
\textit{Observation II\@: Existing search strategies ignore useful feedback and have a low probability of predicting the optimal action, which causes ineffective search bugs.}
To predict the next action (i.e., how to change the model or hyperparameters), existing AutoML engines leverage random walk or statistical models.
Considering the huge search space of AutoML, random methods have low probabilities of choosing an optimal action, which leads to ineffective searches.
Statistical methods (e.g., Bayesian) typically require training tens of thousands of models to fit the statistical models before making good predictions, which is not practical.
Moreover, whether these statistical models can accurately model the AutoML search procedure is questionable.
Existing methods get non-optimal action predictions due to the lack of samples and the doubtable models, which cause ineffective search bugs.
In software engineering research, prior works~\cite{ma2018mode,zhang2021autotrainer} have observed that methods that leverage feedback data (e.g., the provenance of computing, loss values) can achieve better results.
Such feedback is dynamic and context-aware, making them suitable for identifying root causes of observed bugs.

\smallskip\noindent
\textit{Observation III\@: Existing model evaluation component completely ignore the training procedure, which makes effective and efficient search infeasible.} 
In the existing AutoML pipeline, the model evaluation calculates the validation accuracy of the trained model and gives it back to the model generation. 
Validation accuracy measures if a search is successful and finished, and it is essential for AutoML evaluation.
But such a single metric is not sufficient for diagnosing and fixing AutoML bugs.
Adding detailed feedback is critical for automatically repairing AutoML processes.
Moreover, model evaluation always happens at the end stage of a trial, which makes it impossible to identify potential problems and provides timely feedback.
Existing works~\cite{zhang2021autotrainer} have shown that integrating model evaluation in training can help resolve training issues.
The lack of detailed and timely feedback leads to both performance and ineffective search bugs.


\begin{figure}
    \centering     
    \includegraphics[width=0.9\columnwidth]{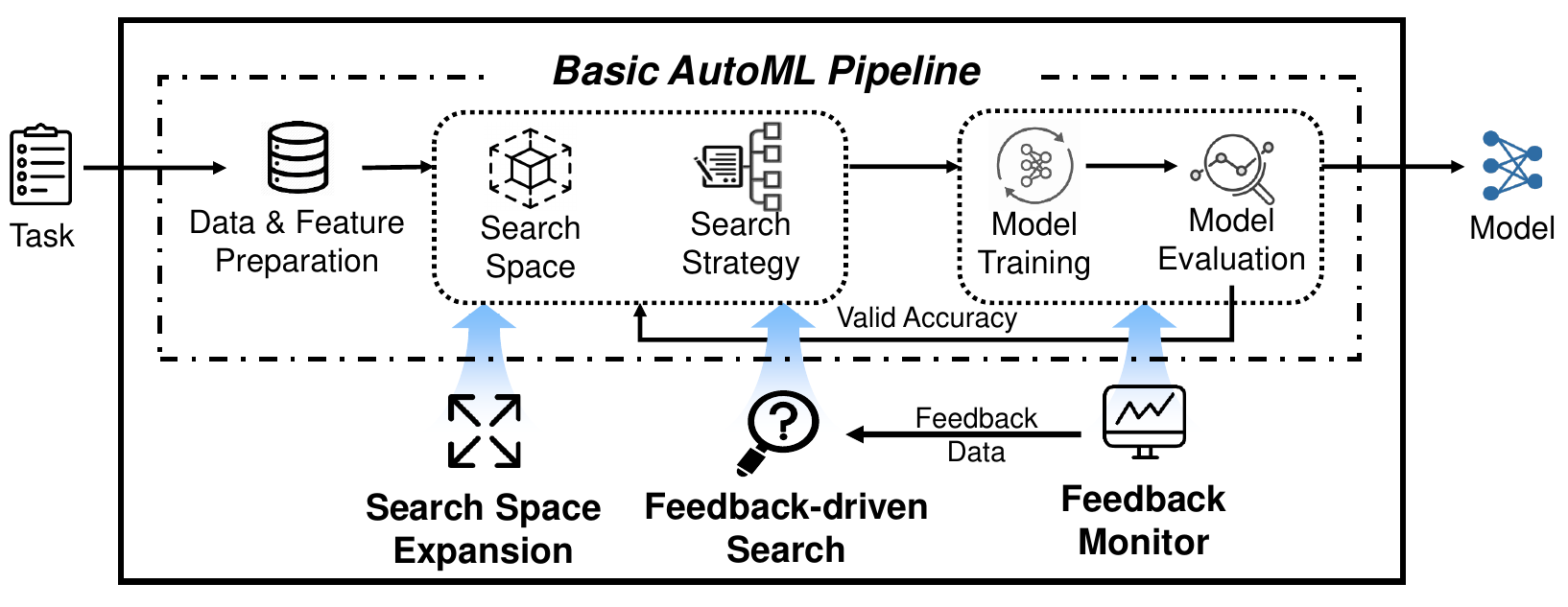}
    \caption{Overarching Design of \sys}\label{fig:overview}
    \vspace{-12pt}
  \end{figure}

\section{Design}\label{sec:design}

In this paper, we present \sys, an automatic debugging and repairing system for the AutoML engines.
It focuses on fixing performance and ineffective search bugs.
\autoref{fig:overview} presents the overview of our design.
It follows the original design of AutoML and adds three new mechanisms: \textit{search space expansion} (enlarging the search space to allow finding optimal models), \textit{feedback-driven search} (a novel search algorithm) and \textit{feedback monitoring} (providing detailed and useful feedback of model training and evaluation, rather than validation accuracy only).
We will introduce search space expansion in \autoref{sec:search}, and discuss the other two components in \autoref{sec:feedback_driven_search}.

\subsection{Search Space Expansion}\label{sec:search}

The search space defines all feasible search actions in the AutoML pipeline in modifying the model architecture and hyperparameters, e.g., changing kernel size, optimizer, architecture type.
As discussed in \autoref{sec:moti}, existing AutoML engines overlook many useful actions.
In \sys, we extend the search space by adding more feasible actions.
\autoref{tab:search_action} shows the description of part of the added search actions. 
The full action table can be found in our repository~\cite{ourrepo}. 
These newly added actions can be categorized as follows:

\makeatletter
\newcommand\newtag[2]{#1\def\@currentlabel{#1}\label{#2}}
\makeatother

\begin{table}[]
    \caption{Newly Added Actions (Partial)}    \label{tab:search_action}
    \centering
    \scriptsize
    \tabcolsep=4pt
    \begin{tabular}{crr}
    \toprule
    Category & \multicolumn{1}{c}{No.} & \multicolumn{1}{c}{Actions Description} \\ \midrule
    \multirow{2}{*}{\begin{tabular}[c]{@{}c@{}}Model\\ Hyperparameters\end{tabular}} & \newtag{1}{que:initializer} & Using other initializer (e.g., He Uniform) as substitution \\
    & \newtag{2}{que:activation} & Using other activation function (e.g., SeLU) as substitution \\ \midrule
    \multirow{3}{*}{\begin{tabular}[c]{@{}c@{}}Optimization\\ Configurations\end{tabular}} & \newtag{3}{que:momentum} & Changing the value of \texttt{momentum} in SGD optimizer\\
    & \newtag{4}{que:end} & Changing the value of \texttt{end learning rate} in AdamW optimizer \\
    & \newtag{5}{que:weight}  & Changing the value of \texttt{weight decay rate} in AdamW optimizer \\ \midrule
    \multirow{5}{*}{\begin{tabular}[c]{@{}c@{}}Fine-tune\\ Strategy\end{tabular}} & \newtag{6}{que:double} & Fine-tuning the pre-trained model with two-step training \\
    & \newtag{7}{que:triple} & Fine-tuning the pre-trained model with three-step training \\
    & \newtag{8}{que:epoch} & Changing the epoch of different steps in training \\
    & \newtag{9}{que:lr} & Changing the learning rate of different steps in training \\
    & \newtag{10}{que:freeze} & Freezing the pre-trained model in training \\
    \bottomrule
    \end{tabular}
    \vspace{-8pt}
\end{table}

\smallskip
\noindent \(\bullet\) 
\textbf{Model hyperparameters}:
Most AutoML engines use fixed model hyperparameters.
For example, all searches in AutoKeras use the ReLU activation function and Xavier initializer.
However, the same activation function and initializer can not work effectively on all models.
Existing works have demonstrated that other activation functions and initializers have the opportunity to improve model performance~\cite{klambauer2017self,he2015delving}.
\sys provides several activation functions and initializers as additional optimization opportunities in the search, e.g., He Uniform~\cite{he2015delving} and the SeLU activation function (Rows~\ref{que:initializer} and~\ref{que:activation} in \autoref{tab:search_action}).

\smallskip
\noindent \(\bullet\) 
\textbf{Optimization configurations}:
The appropriate optimizer configuration and learning rate schedule can play a positive role in model training~\cite{park2020novel,loshchilov2017decoupled,ge2019step}.
Existing AutoML engines use fixed configurations for all searches, which may lead to bugs.
\sys adds several optimization options for different optimizers, for example, the \texttt{momentum} parameter in the SGD optimizer (Row~\ref{que:momentum} in \autoref{tab:search_action}), \texttt{end learning rate} (Row~\ref{que:end}) and \texttt{weight decay rate} (Row~\ref{que:weight}) in the AdamW optimizer referring to the existing works~\cite{qian1999momentum,loshchilov2017decoupled}.

\smallskip
\noindent \(\bullet\) 
\textbf{Fine-tune strategy}: 
Multi-step fine-tune strategy is a commonly used fine-tune strategy for the pre-trained model, and its effectiveness has been verified in existing works~\cite{keras_efficientnet,chollet2021deep,tensorflow_transfer_learning}.
\sys supplements five actions to search space to implement the multi-step fine-tune.
Specifically, \sys adds the two-step and three-step fine-tuning strategies in training, as shown in Row~\ref{que:double} and Row~\ref{que:triple} of \autoref{tab:search_action}.
In addition, three actions are supplemented in the search space to control the transfer learning parameters, i.e., number of epochs, learning rate, and frozen status of the pre-trained model.
These actions are shown from Rows~\ref{que:epoch} to~\ref{que:freeze}.

\subsection{Monitoring \& Feedback-driven Search}\label{sec:feedback_driven_search}
The feedback monitor in \sys collects the feedback data from both model training and evaluation, and the feedback-driven search is a novel search algorithm leveraging such feedback.
Similar to existing AutoML, model search, training, and evaluation are iterative.
\autoref{algo:overview} shows the process.

\renewcommand{\algorithmiccomment}[1]{\ \ \ \ /* \ #1 \ */}
\begin{algorithm}
  \caption{Search \& Repair}\label{algo:overview}
  \footnotesize
  \begin{flalign*}
    \textbf{Input:} \qquad D &- \mbox{the training dataset}; &\\[-0.4em]
    T_{m} &- \mbox{the maximum number of search trials}; &\\[-0.4em]
    S_{t} &- \mbox{the target score in search}; &\\[-0.4em]
    M &- \mbox{the seed model in a trial} &\\[-0.4em]
    \textbf{Output:} \quad M_{c} &- \mbox{the selected model in search}; &\\[-0.4em]
  \end{flalign*}
  \vspace{-18pt}

  \begin{algorithmic}[1]
    

    




  \Procedure{SearchAndRepair}{$D , T_{m}, S_{t}, M$}       
    \State $ t \leftarrow 0 $
      \While{$t < T_{m}$} \label{line:startwhile}

        \Comment{Monitor Feedback and Detect Bugs}
        \State $ F_t, S_t, Bug \leftarrow feedbackMonitor(D,M) $ \label{line:initial}
        
        \Comment{Update Search Result}
        \State $ M_{c}, F_{c}, S_{c} \leftarrow  update(M, F_t, S_t) $\label{line:update}
        
        \Comment{Early Stop}
        \If {$ S_{c} \geq  S_{t} $}
            \State \Return $M_{c}$
            \label{line:early}
        \EndIf

        \If {$ Bug \neq \emptyset $}\label{line:bug} 

            \Comment{Select the Optimal Action Based on Feedback}
            \State $ A \leftarrow  selectAction(F_{c}, M_{c}) $\label{line:action}      
            \State $ M \leftarrow buildModel(M_{c}, A) $\label{line:newmodel}
        \Else

            \Comment{No Bugs, No Fixes}
            \State $ M \leftarrow  searchStrategy(M_{c}, S_{c}) $\label{line:originalsearch}
        \EndIf
        
        \State $ t \leftarrow t + 1 $ \label{line:increase}
      \EndWhile 
      \label{line:endwhile}
  
      \State \Return $M_{c}$
  \EndProcedure

\end{algorithmic}
\end{algorithm}

The input to search and repair components of \sys is the same with existing AutoML: a dataset \(D\), the maximum number of trials \(T_{m}\), the target score \(S_{t}\), and the seed model \(M\).
Lines~\ref{line:startwhile} to~\ref{line:endwhile} of \autoref{algo:overview} show this iterative process.
Firstly, in Line~\ref{line:initial}, the feedback monitor collects feedback data \(F_t\) and model score \(S_t\) from the model training and evaluation in the \(t\)-th trial and detects the performance and ineffective search bugs in the search.
In Line~\ref{line:update}, we update the candidate model \(M_{c}\), corresponding feedback \(F_{c}\) and score \(S_{c}\) based on monitoring results to select the ones with the highest scores.
If we find a satisfactory model before using up all budgets, we exist and return the model (Line~\ref{line:early}).
This is a typical early-stopping mechanism to optimize the search time.
If not, we will continue our search.
In the rest of the trial, we first detect if there exist potential new bugs.
If so (Line~\ref{line:bug}),  we start our feedback-driven search to analyze the feedback data \(F_{c}\) (Line~\ref{line:action}), and then, based on the analysis, the search strategy dynamically selects the optimal action \(A\) to improve the model \(M_{c}\).
Subsequently, model generation will leverage \(A\) and \(M_{c}\) to build the model for the next trial, as shown in Line~\ref{line:newmodel}.
If there is no bug, the pipeline will use its default strategy without repair to continue the search, as shown in Line~\ref{line:originalsearch}.
Each time, we increase the number of trials by one (Line~\ref{line:increase}).
The detection and repair process will continue until the number of trials \(t\) reaches \(T_{m}\) if early-stopping is not triggered.

\subsubsection{Feedback Monitoring}\label{sec:train_monitor}
The feedback monitor has two jobs, i.e., collecting feedback data from AutoML engines and detecting performance and ineffective search bugs.

\smallskip
\noindent \(\bullet\)
\textbf{Feedback Collection.}
Different from the simple feedback of the validation accuracy in the AutoML engines, the feedback monitor in \sys collects various feedback data from the model training and evaluation.
Detailed feedback data can help the feedback-driven search diagnose the bug symptoms and fix the current search.
Referring to the existing work~\cite{zhang2021autotrainer}, the feedback monitor combines the model evaluation into each epoch in training to implement detailed and timely feedback in the pipeline, which helps identify bugs and further repair.
The feedback in \sys include many aspects, which are shown as follows:

\begin{itemize} 
  \item Model architectures and configurations (e.g., model depth, activation
  functions, initializer). 
  \item Optimization method and its parameters (e.g., learning rate). 
  \item Training accuracy, loss value, and other training histories.
  \item Calculated gradients and weights for each neuron in each epoch in training.
  \item Other necessary hyperparameters used in training (e.g., image
        augmentation configurations).
\end{itemize}

\smallskip
\noindent \(\bullet\)
\textbf{Bug Detection.}
To detect the performance and ineffective bugs accurately and in a timely manner, we define and formalize the symptoms of these two bugs.
The specific symptoms of the two bugs are as follows.

\noindent
\underline{\it Performance bug}: When the AutoML engine takes too much time and still cannot reach the target Score \(S_{t}\), we think that the search has the performance bug:
\[
  S_{c}<S_{t} \vee T>T_{r}
\]
where the \(T\) indicates the current time, and the \(T_{r}\) is the time threshold to judge the bug.

\noindent
\underline{\it Ineffective search bug}: When the AutoML engine has no improvement to the model score \(S_{c}\) in the past \(t_{r}\) trials, \sys treats it as an ineffective search bug:
\[
  \forall k \in \left[ t-t_{r},t \right], S_k < S_{c}
\]
where the \(S_k\) indicates the score of the \(k\)-th trial.

\subsubsection{Feedback-driven Search}\label{sec:priority_analyzer}

The feedback-driven search leverages given feedback to select the optimal actions to improve the search effect, corresponding to Line~\ref{line:action} of~\autoref{algo:overview}.

Extracting useful information from the massive feedback data \(F\) is the key to repairing the pipeline.
The feedback-driven search first summarizes \(F\) into four categories, i.e., architectures (\textit{A}), convergence status (\textit{C}), gradients (\textit{G}), and weights (\textit{W}) to describe model training and evaluation from four different angles.
Then, our feedback-driven search can be formalized as finding the optimal action based on given \textit{A-C-G-W} observations.
We solve this problem by calculating the conditional probability, \(\mathbb{P}(Action | A, C, G, W)\) for observed architectures, convergence status, gradients, and weights.


\smallskip
\noindent \underline{\textit{Architectures.}}
The DNN architecture has a significant influence on the model performance.
For example, the residual block in ResNet can effectively speed up the convergence of the training and help deepen the model~\cite{veit2016residual,he2016deep}.
We roughly classify models architectures into four types, \textit{ResNet Architecture} (RA), \textit{EfficientNet Architecture} (EA),
\textit{XceptionNet Architecture} (XA) and \textit{Other Architecture} (OA). 
The first three are widely used DNN architectures~\cite{he2016deep,tan2019efficientnet,chollet2017xception}.
The last one refers to the models without specific architectures.

\smallskip
\noindent \underline{\textit{Convergence Status.}}
We categorize the convergence status of a model based on its validation accuracy and loss value into two types: \textit{Slow Convergence} (SC) and \textit{Normal Convergence} (NC).
SC is a typical root cause of performance bugs.
The model meets the SC condition when the accuracy changes of two contiguous epochs in training are less than 0.01~\cite{zhang2021autotrainer}, and we classify the remaining convergence conditions as NC. 

\smallskip
\noindent \underline{\textit{Gradients.}}
Abnormal gradients in training can implicitly explain failed trials.
We group gradients into four types based on existing works~\cite{hochreiter1991untersuchungen,sussillo2014random,miller2018stable,arnekvist2020effect}, i.e., \textit{Vanishing Gradient} (VG), \textit{Exploding Gradient} (EG), \textit{Dying-ReLU Gradient} (DG), and \textit{Normal Gradient} (NG).
When the ratio of gradients from the input layer to the output layer is smaller than \(1e-3\), \sys considers it as a VG case.
In contrast, if this ratio is over 70, \sys treats it as EG.
DG happens when the number of neurons with zero gradients is 70\% of the total~\cite{zhang2021autotrainer}.
The occurrence of VG, EG, and DG can cause various training problems, including performance and ineffective search bugs.

\smallskip
\noindent \underline{\textit{Weights.}}
Once model weights overflow (or implementation bugs happen), the model performance will be severely impacted.
Therefore, the weight condition is divided into \textit{Explode Weight} (EW) and \textit{Normal Weight} (NW).
If there are \texttt{NaN} values in model weights, the model will be considered as EW. 

\smallskip\noindent 
\textbf{Calculating Conditional Probability.}
Based on summarized observations from feedback data, \sys selects an optimal action that is the most possible to improve the current search by calculating the conditional probability \(\mathbb{P}(Action | A, C, G, W)\).
Although the existing studies have mentioned that several actions can improve the model performance in some situations~\cite{klambauer2017self,lu2019dying,goodfellow2016deep}, it is still difficult to directly measure the actual effect of each action under different conditions.
To solve this problem, we construct large-scale experiments to evaluate the search priority of each action under different conditions in the feedback-driven search.
Then, we sort all calculated probabilities and based on these probabilities, we can have a prioritized action list in search.
In each trial, we choose the action with the highest probability (under observed A-C-G-W) to help generate new models.
Detailed experiments and results, search priorities, and an example can be found in our repository~\cite{ourrepo}.
Our anonymized supplementary material contains an extended version of the paper that also describes these details.
\section{Evaluation}\label{sec:eval_set}

In this section, we aim to answer the following research questions.

\noindent \textbf{RQ1:}
How effective is \sys in fixing AutoML bugs?

\noindent \textbf{RQ2:}
How efficient is \sys?

\noindent \textbf{RQ3:}
How effective is each design in \sys?

\noindent \textbf{RQ4:}
What is the impact of different action priorities in repairing?

\subsection{Setup}\label{s:setup}

\noindent
{\bf Datasets:}
We perform our experiments on four popular datasets: CIFAR-100~\cite{cifar100_dataset}, Food-101~\cite{bossard2014food}, Stanford Cars~\cite{krause20133d}, and TinyImageNet~\cite{le2015tiny}.
CIFAR-100 is a widely-used colored image dataset used for object recognition and contains 100 categories.
Food-101 is an image classification dataset with 101 food categories and a total of 101k images.
In the experiments, we resize all images in the Food-101 dataset to the size of \(300\ast300\).
Stanford Cars is an image dataset with 16,185 images of 196 classes of cars.
These images of cars are resized to \(360\ast240\).
TinyImageNet dataset consists of 100k colored images of 200 classes, and all images have been downsized to \(64\ast64\).
For a fair comparison, all strategies in experiments use the same data preprocessing and training procedures.

\noindent
{\bf Baseline and Metric:}
We use three implemented strategies in AutoKeras for comparison, i.e., Greedy, Bayesian, and Hyperband.
Detail descriptions about each strategy are shown in~\autoref{sec:automl}.
In the experiments, we use \textit{Validation Accuracy} as the \textit{Score} to evaluate the model performance in searches.
We use \textit{Hours} to refer to the \textit{GPU Hours} in searches.
To ensure that the models can be fully trained, the maximum epoch in training is set to 200.
In addition, we enable the early stopping setting to speed up the training process.
It is also the default setting in the AutoKeras engine to terminate each training process in advance and improve the search efficiency.
If not specified, the threshold parameters \(T_{r}\) and \(t_{r}\) of the feedback monitor in \sys are set to 0 to intuitively show the effectiveness of \sys in repairing the bugs.

\noindent
{\bf Software and Hardware:}
The prototype of \sys implements on the top AutoKeras 1.0.12~\cite{AutoKeras} and TensorFlow 2.4.3~\cite{abadi2016tensorflow}.
If not specified, all searches in the experiments of this section start with the same code recommended by AutoKeras~\cite{AutoKeras}.
And the training process uses 32 as the batch size.
All experiments are conducted on a server with Intel(R) Xeon E5-2620 2.1GHz 8-core processors, 130 GB of RAM, and an NVIDIA RTX 3090 GPU running Ubuntu 20.04 as the operating system.

\subsection{Effectiveness of \sys}\label{s:effective}

\noindent
{\bf Experiment Design:}
To evaluate the effectiveness of \sys in fixing the AutoML ineffective search bug and performance bug, we conduct comparative experiments with a total of eighteen searches on four datasets with each search strategy (i.e., three baseline strategies in AutoKeras and the repaired search in \sys).
These searches are used to observe whether \sys can effectively repair the bugs in the AutoML pipeline and guide the search to perform more effectively.
To have a fair comparison,
we randomly generate the initial model with AutoKeras and ensure that each search strategy starts with the same initial model.
Each search runs in the experiment for at least 24 hours and 25 trials, and we set the target score as 70\% accuracy in these searches.
To evaluate the effectiveness of \sys in fixing the ineffective search bug, we record the search results and accuracy improvements of all the strategies in 25 trials.
We also record the time cost of each search strategy reaching the search accuracy target and the highest score within 24 hours in the whole search process to evaluate whether the performance bug has been repaired by \sys effectively.
All other environment configurations are set as described in~\autoref{s:setup}.

\begin{table*}[]
	\centering
	\scriptsize
	\caption{Overall Search Results on Four Datasets}
	\label{t:effective_result}
	\tabcolsep=5pt
	\begin{tabular}{crrrrrrrrrrrrrrrrrr}
	\toprule
	& \multicolumn{1}{c}{} & \multicolumn{1}{c}{} & \multicolumn{4}{c}{Best Score in 25 Trials (\%)} & \multicolumn{4}{c}{Score Improvement (\%)} &
	\multicolumn{4}{c}{Time to Reach Target (h)} & \multicolumn{4}{c}{Best Score in 24 Hours (\%)} \\ \cmidrule(r){4-7} \cmidrule(r){8-11}
	\cmidrule(r){12-15} \cmidrule(r){16-19}  
	\multirow{-2}{*}{Dataset} & \multicolumn{1}{c}{\multirow{-2}{*}{No.}} & \multicolumn{1}{c}{\multirow{-2}{*}{\begin{tabular}[c]{@{}c@{}}Initial\\
	Score(\%)\end{tabular}}} & \multicolumn{1}{c}{AKG} & \multicolumn{1}{c}{AKB} & \multicolumn{1}{c}{AKH} & \multicolumn{1}{c}{\sys} &
	\multicolumn{1}{c}{AKG} & \multicolumn{1}{c}{AKB} & \multicolumn{1}{c}{AKH} & \multicolumn{1}{c}{\sys} & \multicolumn{1}{c}{AKG} &
	\multicolumn{1}{c}{AKB} & \multicolumn{1}{c}{AKH} & \multicolumn{1}{c}{\sys} & \multicolumn{1}{c}{AKG} & \multicolumn{1}{c}{AKB} &
	\multicolumn{1}{c}{AKH} & \multicolumn{1}{c}{\sys} \\ \midrule
	& 1 & 10.59 & 63.04 & 65.77 & 78.78 & \cellcolor{blue!20}83.32 & 52.45 & 55.18 & 68.19 & \cellcolor{blue!20}72.72 & - & - & 1.63 & \cellcolor{blue!20}0.92 & 63.04 & 52.51 & 79.19 & \cellcolor{blue!20}83.32 \\
	& 2 & 20.74 & 65.38 & 78.88 & 36.35 & \cellcolor{blue!20}85.85 & 44.64 & 58.15 & 15.61 & \cellcolor{blue!20}65.12 & - & 47.52 & 2.45 & \cellcolor{blue!20}2.34 & 59.51 & 45.29 & 81.57 & \cellcolor{blue!20}83.55 \\
	& 3 & 2.58 & 34.23 & 67.79 & 68.21 & \cellcolor{blue!20}82.66 & 31.65 & 65.21 & 65.63 & \cellcolor{blue!20}80.09 & 18.32 & - & 5.45 & \cellcolor{blue!20}1.80 & 77.79 & 67.79 & 80.21 & \cellcolor{blue!20}82.66 \\
	& 4 & 8.19 & 37.63 & 64.49 & 73.15 & \cellcolor{blue!20}82.62 & 29.44 & 56.30 & 64.97 & \cellcolor{blue!20}74.44 & - & 17.97 & 1.36 & \cellcolor{blue!20}0.53 & 42.57 & 76.52 & 78.89 & \cellcolor{blue!20}82.62 \\
	& 5 & 4.81 & 64.82 & 46.92 & 54.30 & \cellcolor{blue!20}83.03 & 60.01 & 42.11 & 49.50 & \cellcolor{blue!20}78.22 & - & - & 20.49 & \cellcolor{blue!20}0.53 & 63.61 & 33.03 & 77.01 & \cellcolor{blue!20}83.03 \\
   \multirow{-6}{*}{CIFAR-100} & Avg. & 9.38 & 53.02 & 64.77 & 62.16 & \cellcolor{blue!20}83.50 & 43.64 & 55.39 & 52.78 & \cellcolor{blue!20}74.12 & - & - & 6.28 & \cellcolor{blue!20}1.22 & 61.30 & 55.03 & 79.37 & \cellcolor{blue!20}83.04 \\ \midrule
	& 6 & 8.37 & 72.67 & 57.47 & 51.59 & \cellcolor{blue!20}83.13 & 64.31 & 49.10 & 43.22 & \cellcolor{blue!20}74.77 & 40.35 & - & - & \cellcolor{blue!20}21.66 & 61.68 & 57.47 & 59.42 & \cellcolor{blue!20}71.56 \\
	& 7 & 2.30 & 61.44 & 41.37 & 50.16 & \cellcolor{blue!20}78.23 & 59.14 & 39.07 & 47.86 & \cellcolor{blue!20}75.93 & - & - & - & \cellcolor{blue!20}45.04 & 13.57 & 31.93 & 58.74 & \cellcolor{blue!20}65.74 \\
	& 8 & 8.48 & 68.80 & 75.72 & 55.61 & \cellcolor{blue!20}82.62 & 60.32 & 67.25 & 47.13 & \cellcolor{blue!20}74.14 & - & 82.24 & - & \cellcolor{blue!20}13.34 & 13.54 & 35.94 & 55.61 & \cellcolor{blue!20}72.03 \\
   \multirow{-4}{*}{Food-101} & Avg. & 6.38 & 67.64 & 58.19 & 52.45 & \cellcolor{blue!20}81.33 & 61.26 & 51.80 & 46.07 & \cellcolor{blue!20}74.95 & - & - & - & \cellcolor{blue!20}26.68 & 29.60 & 41.78 & 57.92 & \cellcolor{blue!20}69.78 \\ \midrule
	& 9 & 0.74 & 24.75 & 79.64 & 48.33 & \cellcolor{blue!20}85.02 & 24.01 & 78.90 & 47.59 & \cellcolor{blue!20}84.28 & - & 9.48 & - & \cellcolor{blue!20}4.03 & 30.76 & 77.60 & 55.57 & \cellcolor{blue!20}83.79 \\
	& 10 & 0.56 & 42.70 & 81.99 & 24.44 & \cellcolor{blue!20}83.66 & 42.14 & 81.44 & 23.89 & \cellcolor{blue!20}83.11 & 16.87 & 8.38 & - & \cellcolor{blue!20}7.85 & 71.53 & 81.99 & 46.66 & \cellcolor{blue!20}83.35 \\
	& 11 & 0.80 & 60.33 & 16.40 & 20.17 & \cellcolor{blue!20}83.48 & 59.53 & 15.59 & 19.37 & \cellcolor{blue!20}82.67 & - & - & - & \cellcolor{blue!20}17.57 & 7.61 & 16.40 & 40.72 & \cellcolor{blue!20}70.61 \\
	& 12 & 0.62 & 44.06 & 46.23 & 32.86 & \cellcolor{blue!20}83.42 & 43.44 & 45.61 & 32.24 & \cellcolor{blue!20}82.80 & - & - & - & \cellcolor{blue!20}7.06 & 14.79 & 46.23 & 58.17 & \cellcolor{blue!20}83.42 \\
	& 13 & 0.68 & 39.42 & 8.29 & 16.58 & \cellcolor{blue!20}84.10 & 38.74 & 7.61 & 15.90 & \cellcolor{blue!20}83.42 & 19.57 & - & - & \cellcolor{blue!20}5.33 & 81.31 & 8.29 & 43.01 & \cellcolor{blue!20}84.10 \\
   \multirow{-6}{*}{\begin{tabular}[c]{@{}c@{}}Stanford\\ Cars\end{tabular}} & Avg. & 0.68 & 42.25 & 46.51 & 28.48 & \cellcolor{blue!20}83.94 & 41.57 & 45.83 & 27.80 & \cellcolor{blue!20}83.25 & - & - & - & \cellcolor{blue!20}8.37 & 41.20 & 46.10 & 48.82 & \cellcolor{blue!20}81.05 \\ \midrule
	& 14 & 21.57 & 29.34 & 21.57 & 71.51 & \cellcolor{blue!20}83.80 & 10.45 & 2.68 & 52.62 & \cellcolor{blue!20}64.92 & - & - & \cellcolor{blue!20}1.71 & 2.60 & 27.38 & 21.57 & 76.98 & \cellcolor{blue!20}77.67 \\
	& 15 & 9.44 & 62.86 & 42.29 & 44.70 & \cellcolor{blue!20}79.90 & 53.43 & 32.86 & 35.27 & \cellcolor{blue!20}70.46 & - & - & 4.17 & \cellcolor{blue!20}2.85 & 62.78 & 42.29 & \cellcolor{blue!20}80.26 & 76.66 \\
	& 16 & 4.84 & 47.44 & 66.39 & 61.04 & \cellcolor{blue!20}83.95 & 42.60 & 61.56 & 56.20 & \cellcolor{blue!20}79.12 & - & - & 10.00 & \cellcolor{blue!20}2.54 & 42.13 & 41.70 & \cellcolor{blue!20}80.09 & 77.02 \\
	& 17 & 7.68 & 37.89 & 46.74 & 53.59 & \cellcolor{blue!20}83.60 & 30.21 & 39.06 & 45.91 & \cellcolor{blue!20}75.92 & - & - & 13.50 & \cellcolor{blue!20}4.41 & 35.24 & 46.74 & 76.61 & \cellcolor{blue!20}77.22 \\
	& 18 & 8.54 & 63.20 & 75.89 & 67.14 & \cellcolor{blue!20}85.41 & 54.66 & 67.35 & 58.60 & \cellcolor{blue!20}76.87 & - & 39.35 & 7.63 & \cellcolor{blue!20}2.44 & 63.20 & 38.28 & 77.10 & \cellcolor{blue!20}85.41 \\
   \multirow{-6}{*}{\begin{tabular}[c]{@{}c@{}}Tiny\\ ImageNet\end{tabular}} & Avg. & 10.41 & 48.14 & 50.57 & 59.59 & \cellcolor{blue!20}83.33 & 38.27 & 40.70 & 49.72 & \cellcolor{blue!20}73.46 & - & - & 7.40 & \cellcolor{blue!20}2.97 & 46.15 & 38.12 & 78.21 & \cellcolor{blue!20}78.79 \\ \midrule
   \multicolumn{2}{c}{Avg.} & 6.75 & 51.11 & 54.66 & 50.47 & \cellcolor{blue!20}83.21 & 44.51 & 48.06 & 43.87 & \cellcolor{blue!20}76.61 & - & - & - & \cellcolor{blue!20}7.94 & 46.22 & 45.64 & 66.99 & \cellcolor{blue!20}79.10 \\ \bottomrule   
	\end{tabular}
  \end{table*}


\begin{figure}[]
	\centering
	\footnotesize
	\begin{subfigure}[t]{0.47\columnwidth}
		\centering     
		\footnotesize
		\includegraphics[width=\columnwidth]{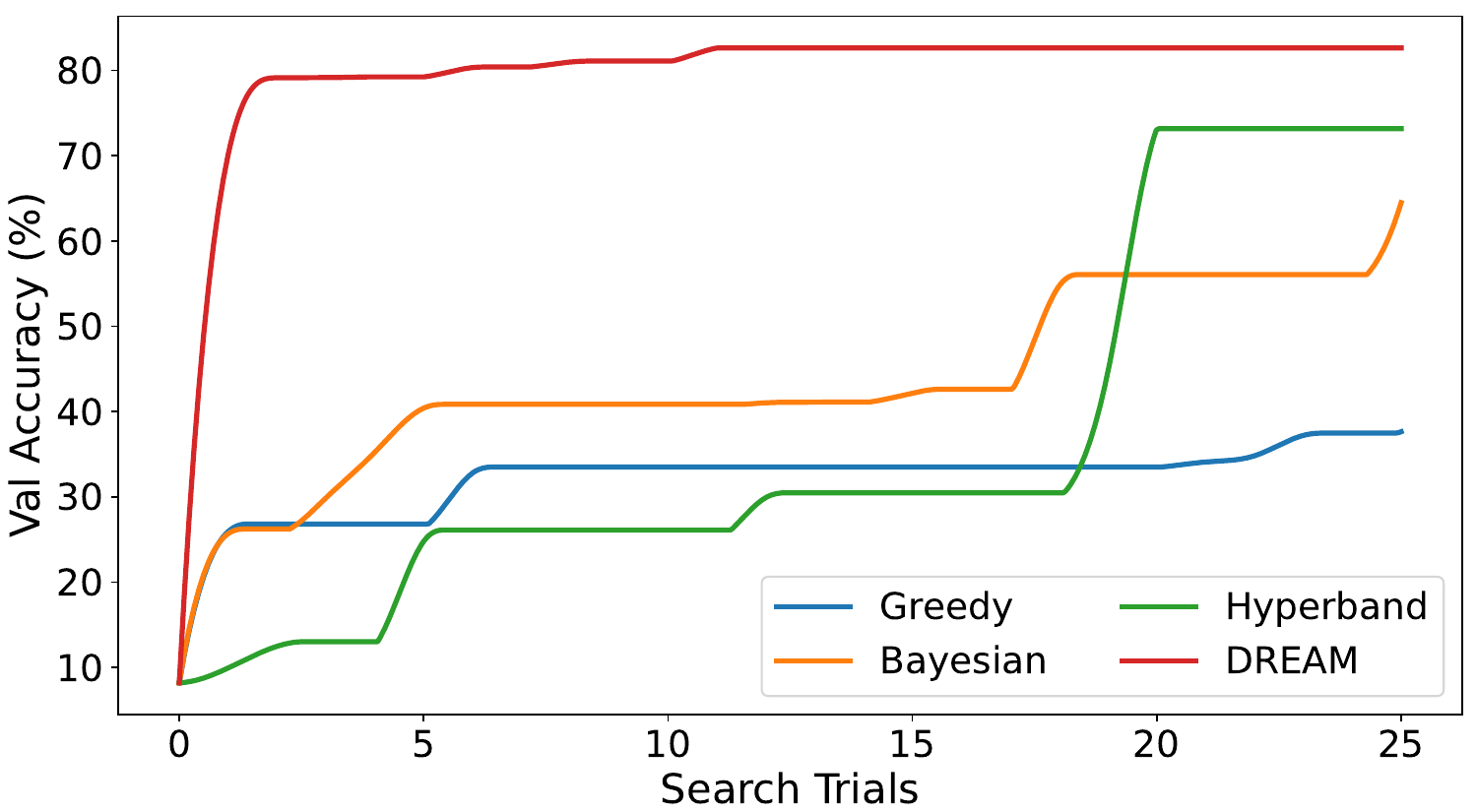}
		\caption{No.4 Search on CIFAR-100 Dataset}
		\label{fig:rq1_c100}
	\end{subfigure}
	\hfill
	\begin{subfigure}[t]{0.47\columnwidth}
		\centering     
		\footnotesize
		\includegraphics[width=\columnwidth]{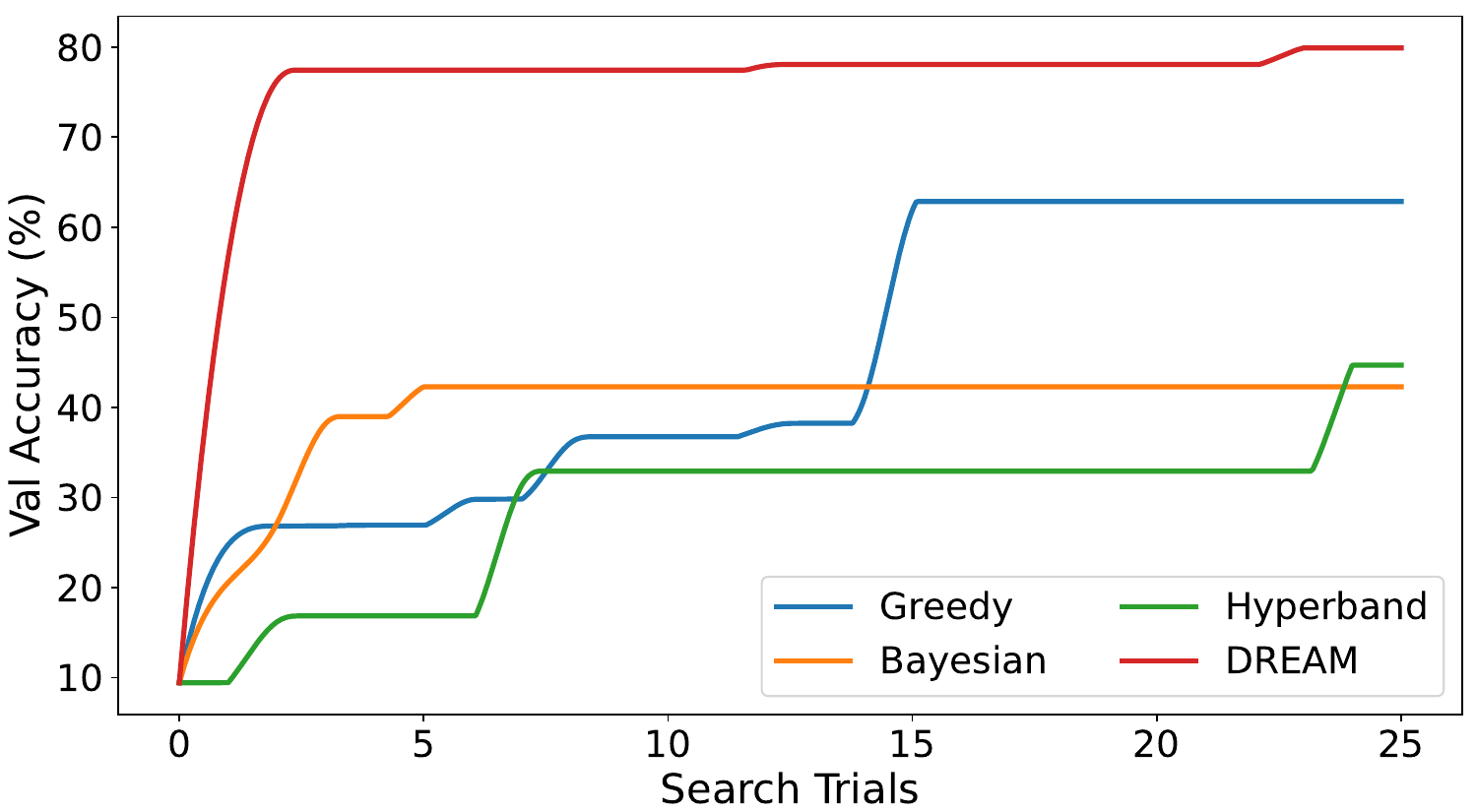}
		\caption{No.15 Search on Tiny ImageNet Dataset}
		\label{fig:rq1_tiny}
	\end{subfigure}
	\caption{Evaluation Results in 25 Trials on CIFAR-100 and Tiny ImageNet Datasets}
	\label{fig:rq1_results}
\end{figure}

\begin{figure}[]
	\centering
	\footnotesize
	\begin{subfigure}[t]{0.47\columnwidth}
		\centering     
		\footnotesize
		\includegraphics[width=\columnwidth]{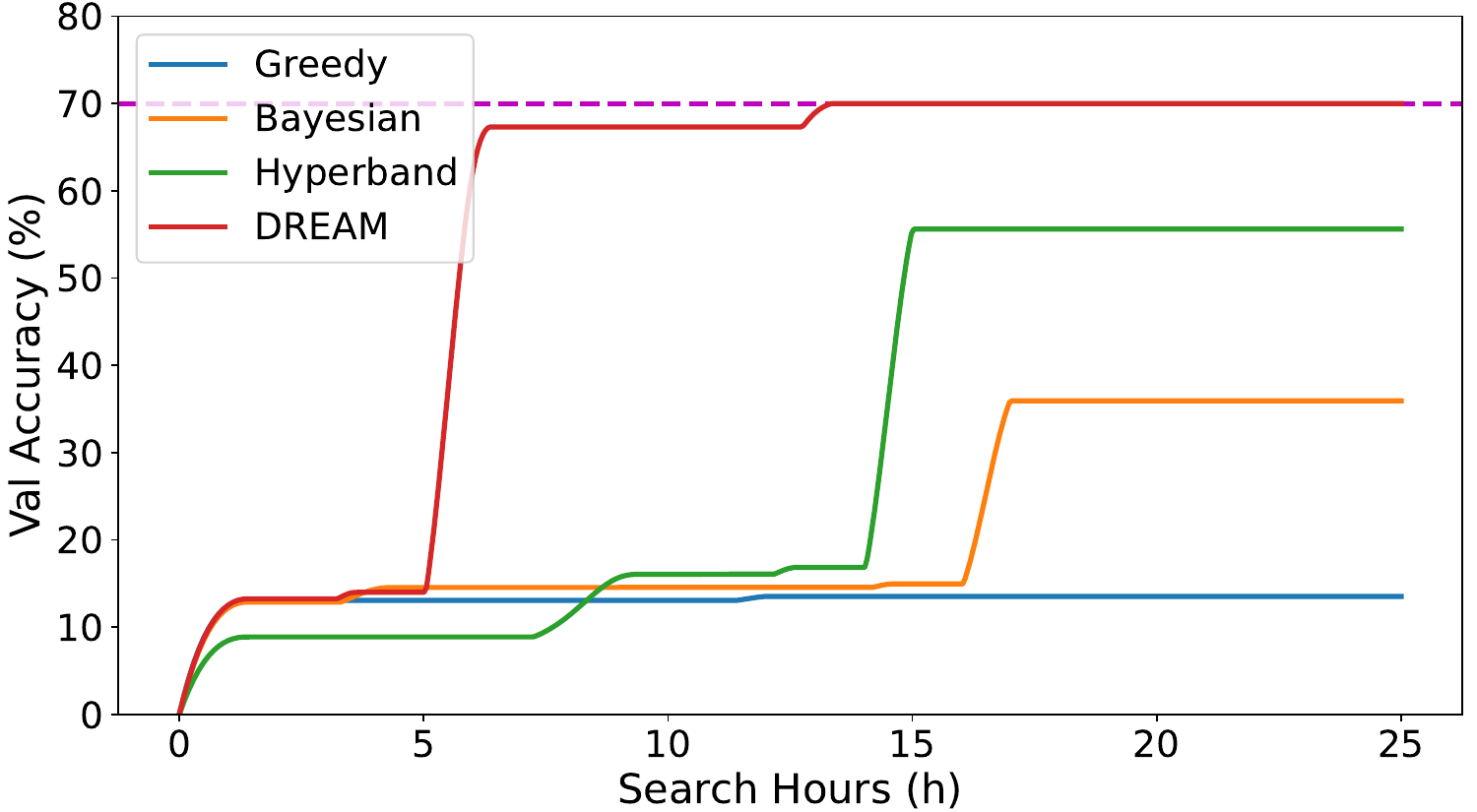}
		\caption{No.8 Search on Food-101 Dataset}
		\label{fig:rq2_f101}
	\end{subfigure}
	\hfill
	\begin{subfigure}[t]{0.47\columnwidth}
		\centering     
		\footnotesize
		\includegraphics[width=\columnwidth]{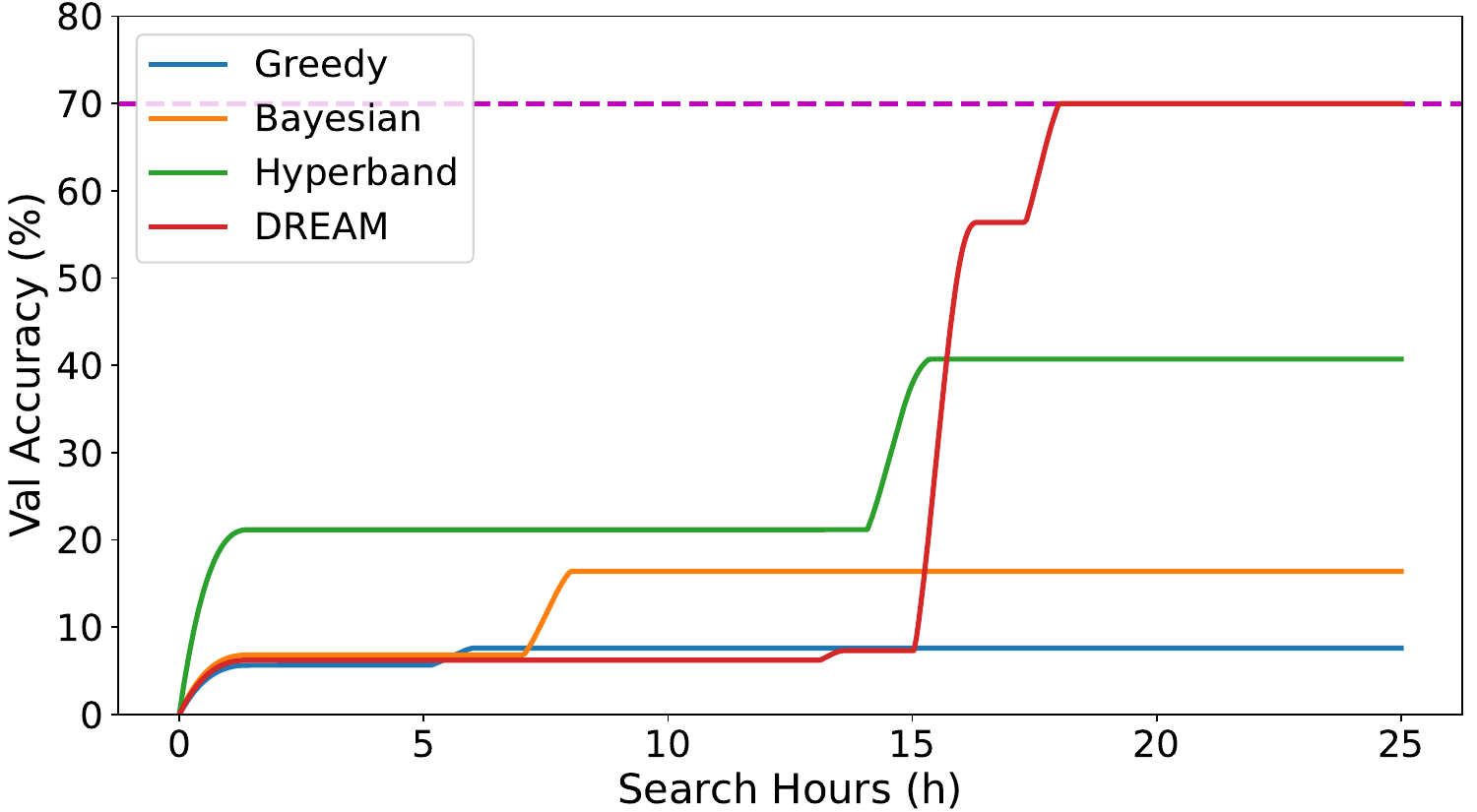}
		\caption{No.11 Search on Standford Cars Dataset}
		\label{fig:rq2_car196}
	\end{subfigure}
	\caption{The Search Results in 24 Hours on Food-101 and Stanford Cars Datasets}
	\label{fig:rq2_results}
\end{figure}

\begin{figure}[]
	\centering
	\footnotesize
	\begin{subfigure}[t]{0.32\columnwidth}
		\centering     
		\footnotesize
		\includegraphics[width=\columnwidth]{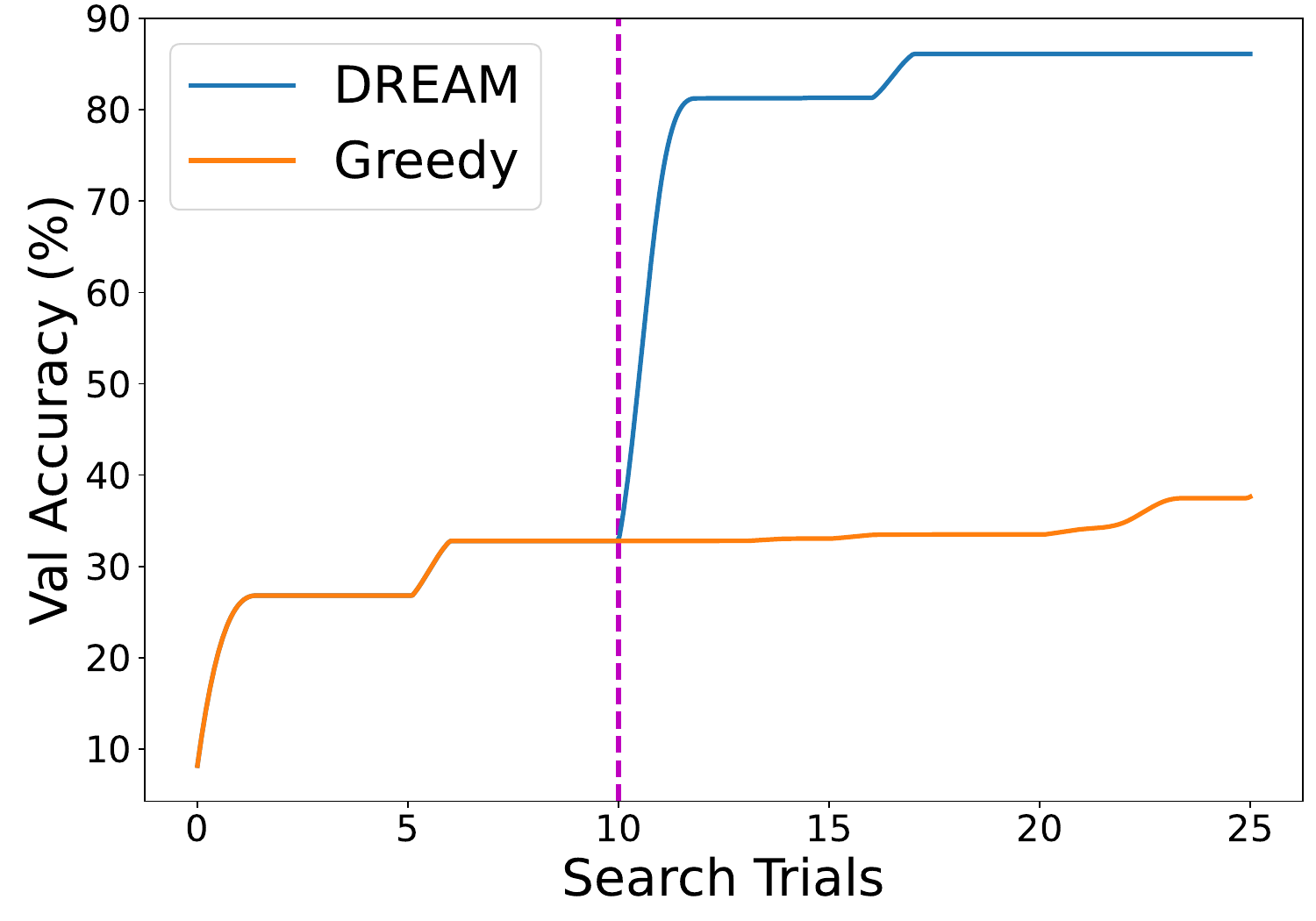}
		\caption{Repair on Greedy}
		\label{fig:rq1_greedy}
	\end{subfigure}
	\hfill
	\begin{subfigure}[t]{0.32\columnwidth}
		\centering     
		\footnotesize
		\includegraphics[width=\columnwidth]{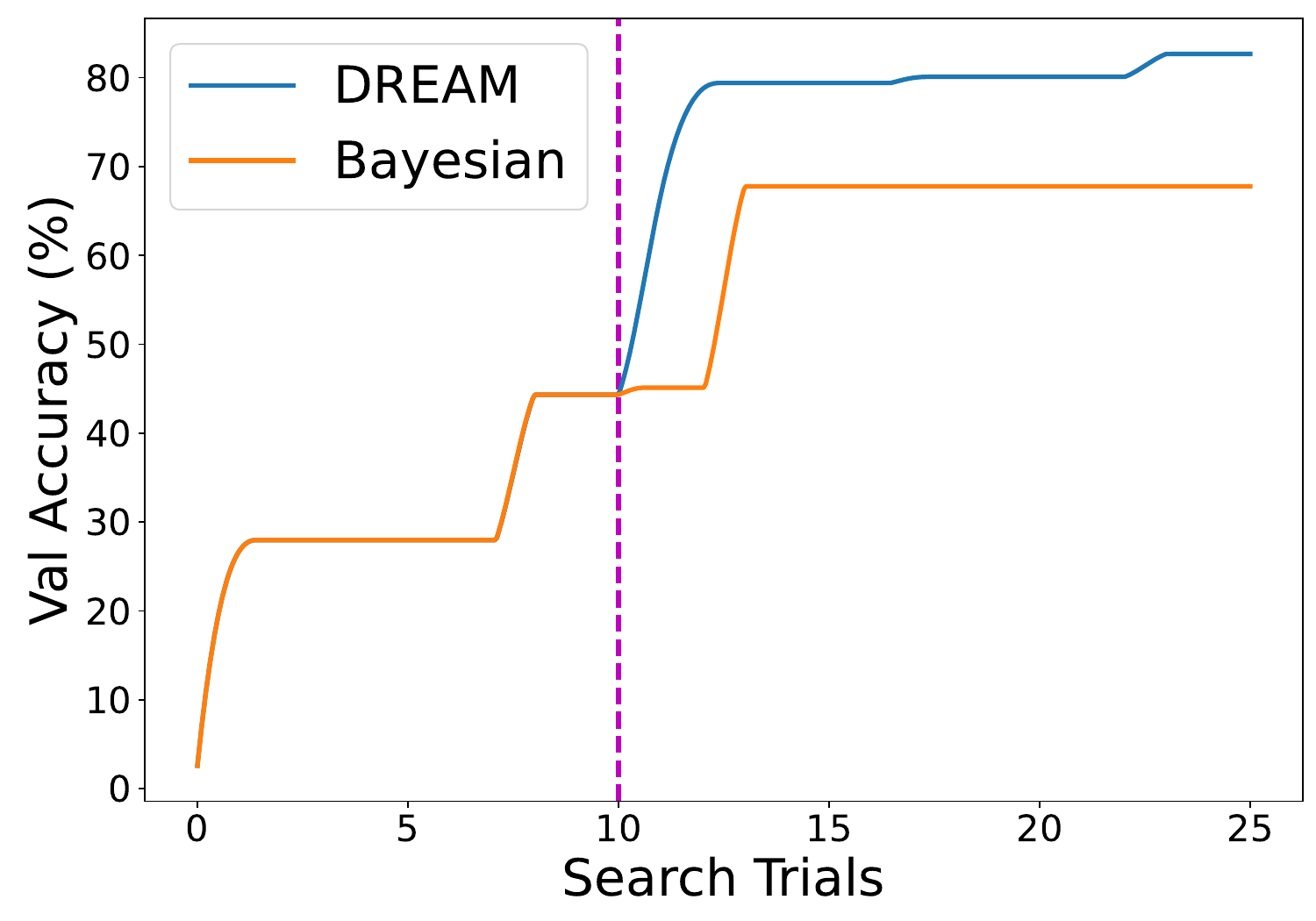}
		\caption{Repair on Bayesian}
		\label{fig:rq1_bayesian}
	\end{subfigure}
	\hfill
	\begin{subfigure}[t]{0.32\columnwidth}
		\centering     
		\footnotesize
		\includegraphics[width=\columnwidth]{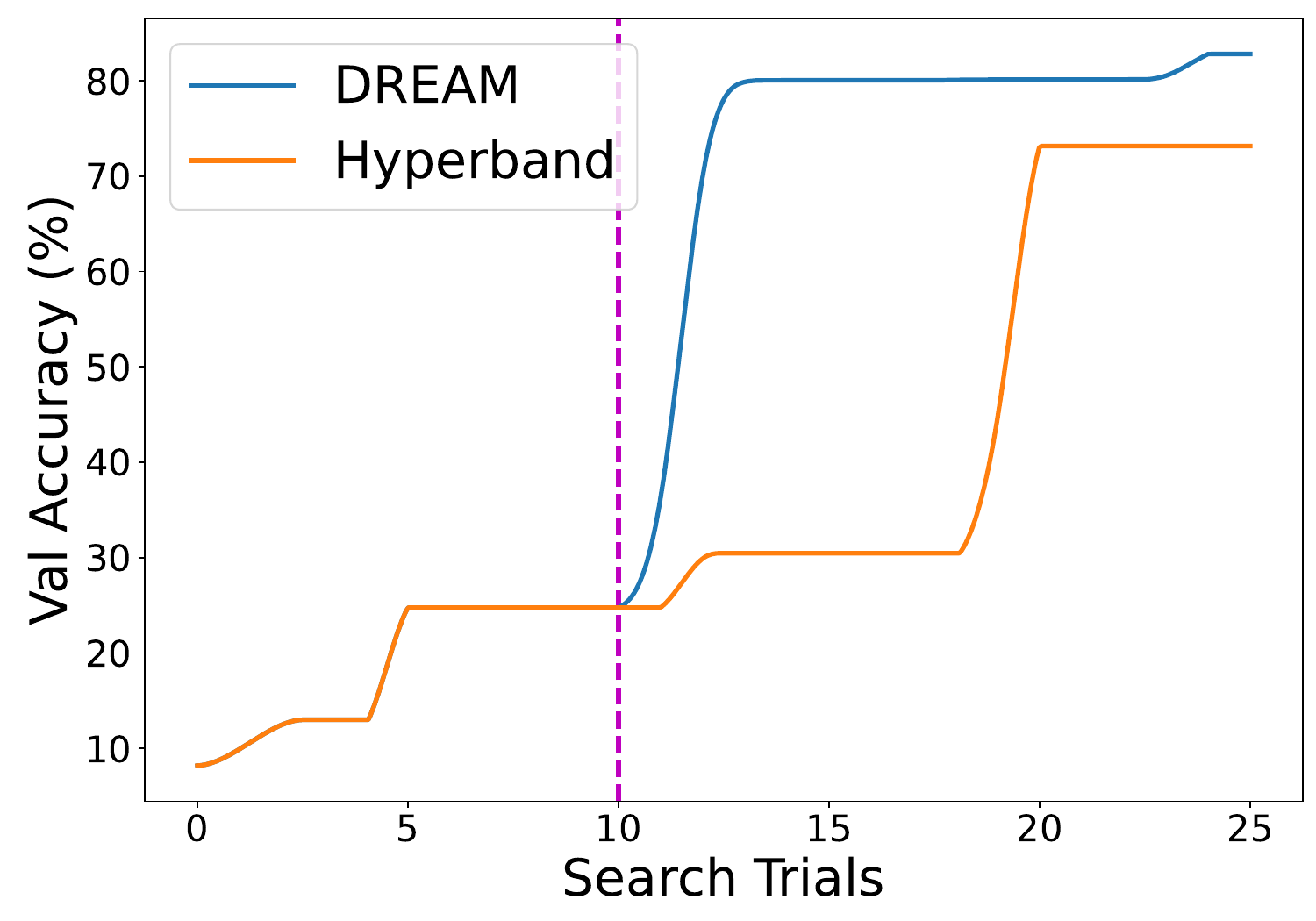}
		\caption{Repair on Hyperband}
		\label{fig:rq1_hyperband}
	\end{subfigure}
	\caption{The Results of \sys Repair the Search After 10 Trials}
	\label{fig:rq1_repair_results}
\end{figure}

\smallskip
\noindent
{\bf Results:} \autoref{t:effective_result} shows the detailed results of the effectiveness in repairing bugs on eighteen searches with 25 search trials and 24 hours.
The first column shows the four datasets, and the second column lists the serial numbers of the searches so that we can specify the searches and analyze them.
The third column shows the score of the initial model for all four search strategies.
The column ``Best Score in 25 Trials'' denotes the best score each strategy achieves in 25 search trials, and the column ``Score Improvement'' shows the improvement of the best score in the previous column compared with the score of the initial model.
The column ``Time to Reach Target'' signifies the time cost of each search strategy to achieve the target score of 70\%.
The `-' in this column represents that the corresponding search strategy fails to reach the target score in the entire search of no less than 24 hours and 25 trials.
When we calculate the average time to reach the target score, we only consider the cases where all searches have reached the target accuracy.
The ``Best Score in 24 Hours'' in this table indicates the best search score of each strategy within 24 hours.
In addition, the cells in blue separately correspond to the searches with the highest score in 25 trials, the maximum improvement, the least time cost to reach the target, and the highest score in 24 hours.
The columns ``AKG'', ``AKB'', and ``AKH'' refer to the abbreviation of the search strategies implemented in the AutoKeras pipeline, i.e., Greedy, Bayesian, and Hyperband.
The column \sys shows the results of the searches repaired by our system.
In addition, to demonstrate the effectiveness of \sys in repairing the performance bugs, \autoref{fig:rq2_results} displays the comparison chart between the results of four strategies in No.8 and No.11 searches within 24 hours.
The X-axis in these figures is the search time, and the Y-axis is the best score that the search achieved.
We also show the comparison between four strategies of No.4 and No.15 searches in 25 trials in~\autoref{fig:rq1_results} to illustrate the repair effect in the ineffective bug of the AutoML pipeline.
The X-axis in the figure is the number of the searched trials, and the Y-axis is the best score that the search achieved in 25 trials.
More details of the experiment results are shown in our repository~\cite{ourrepo}.

\smallskip
\noindent
{\bf Analysis:}
The experiment results illustrate the effectiveness of \sys in repairing the ineffective search bug and the performance bug of the AutoML pipeline.

From~\autoref{t:effective_result}, we can observe that the repaired search processes of \sys can search effectively on the datasets and obtain outstanding performance in limited time and trials.
The columns of ``Best Score in 25 Trial'' and ``Score Improvement'' show that the repaired searches of \sys achieve an average score of 83.21\% on four datasets within 25 trials, which is significantly better than the score of other baselines (i.e., 51.11\%, 54.66\%, 50.47\%).
On the CIFAR-100 dataset, the search repaired by \sys has achieved an average score of 83.50\%, which is an improvement of 74.12\% (i.e., 7.90 times improvement) compared with the initial score of 9.38\%.
The search result in \sys is better than the best scores of 53.02\%, 64.77\%, 62.16\% of the three baselines in AutoKeras.
On the Food-101 dataset and the TinyImageNet dataset, the improvement of the search process in \sys is 11.74 and 7.00 times the average score of the initial models.
This advantage in search effectiveness is even more pronounced in the Stanford Cars dataset, which contains 196 classes.
In 25 search trials, \sys achieves an average best score of 83.94\% on this dataset, higher than the scores of 42.25\%, 46.51\%, 28.48\% for the other three strategies.
Compared with the initial model accuracy of 0.68\%, the search result of \sys improved by 122.31 times.
The advantage of the search effect means that \sys can effectively fix the ineffective search bug.
Therefore, the repaired search process can continuously improve the accuracy in the limited trials, eventually reaching a score far exceeding the score in the original pipeline.
Moreover, \autoref{fig:rq1_results} shows the details of No.4 and No.15 searches on the CIFAR-100 and TinyImageNet datasets, which also illustrate the effectiveness of \sys in repairing the ineffective search bug.
We can observe that the search process repaired by \sys can conduct an effective model search on the given dataset.
It can quickly find a model with high accuracy through several trials and continue to improve in the subsequent search.
Take~\autoref{fig:rq1_tiny} as an example, the search repaired with \sys achieves a score of about 76\% in the second trial.
Then it continues to improve the performance in the 5th, 12th, and 23rd trials and finally reaches the score of 79.90\%.
By contrast, the best search strategy implemented in AutoKeras still suffers from the ineffective search bug.
It only achieved 62.86\% in all 25 trials, and after the 16th search, it couldn't improve anymore.

In order to further present the effectiveness of \sys in repairing bugs, we repair some searches that are suffering from the ineffective search bug.
These searches are guided by AutoKeras search strategies, but their performance cannot improve significantly after 10 search trials.
\sys takes their 10th trials as the starting points and then repairs and conducts the searches for another 15 trials.
\autoref{fig:rq1_repair_results} shows a comparison of the searches fixed by \sys and the original searches on CIFAR-100 dataset, where the purple dotted line represents the starting point of the repair searches of \sys.
We can observe that the effectiveness of the repaired search is significantly improved. It can quickly improve the accuracy over 75\% within a few trials, far exceeding other search strategies in the AutoKeras pipeline.

The last two columns in~\autoref{t:effective_result}
shows that all searches repaired by \sys achieve the target accuracy of 70\% in the search process on four datasets.
These searches in \sys obtain an average score of 79.10\% in 24 hours and take an average of 7.94 hours to reach the target accuracy on four datasets.
However, only 20/54 searches performed by the three search strategies in AutoKeras achieve the target accuracy, which suffers from the performance bug.
Among them, the Hyperband search strategy with the best search performance only achieves the target score in 10/18 searches, and the time cost is much higher than that of the search in \sys.
In addition, the accuracy of the search strategies in AutoKeras within 24 hours are 46.22\%, 45.64\%, 66.99\%, respectively, which are lower than the search results of \sys.
On the CIFAR-100 dataset, the search efficiency of AutoKeras strategies is relatively acceptable.
8/15 searches on this dataset have achieved the target accuracy of 70\%.
However, on the three large-scale datasets of Food-101, Stanford Cars, and TinyImageNet, the search performance of AutoKeras degrades significantly.
There are 2, 4, and 6 searches that reach the target score on the three datasets, respectively.
These experimental results demonstrate that \sys can effectively address the performance bug in the AutoML pipeline.
The fixed searches in \sys can achieve good search results in a limited time and quickly reach the search targets.
\autoref{fig:rq2_results} shows the details of No.8 and No.11 searches on the Food-101 and Stanford Cars datasets, which also illustrate the effectiveness of repairing the performance bug in \sys intuitively.
Take~\autoref{fig:rq2_f101} for example, we can notice that the search of \sys reaches an accuracy of nearly 70\% within 6 hours after starting searching, and further improves and reaches the target score in 13.34 hours.
Hyperband strategy with the best performance of the AutoKeras search strategies only achieves an accuracy of 55.61\% in about 15 hours and ultimately fails to meet the search target.
Bayesian search strategy, which solves the model searching as ML problem and learns from the search histories, only obtains an accuracy of 35.94\% in 24 hours and finally spends 82.24 hours to reach the target score.
Compared with the Bayesian search strategy, the efficiency of the repaired search of \sys improves by 5.16 times, and \sys saves 24.14 kWh electricity and emits 10.43 kg less CO2 and carbon footprints in this single search.
The above results prove the effectiveness of \sys in repairing the performance bug, and the searches in \sys have significantly higher search performance and search efficiency.
\subsection{Efficiency of \sys}\label{s:efficiency}

\begin{figure}
	\centering
	\includegraphics[width=0.9\linewidth]{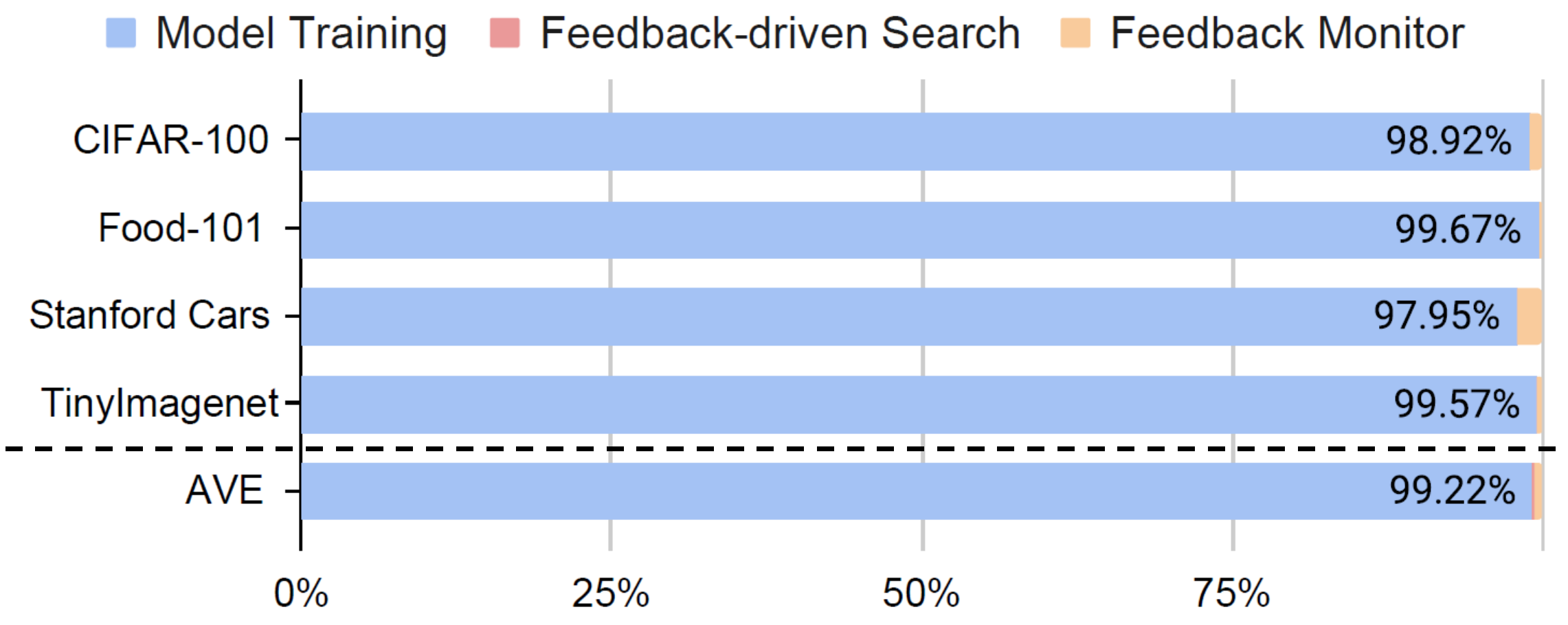}
	\caption{Time Overhead Distribution of \sys in the Searches on Four Datasets}
	\label{fig:timeoverhead}
\end{figure}

\noindent
{\bf Experiment Design and Results:}
To evaluate the efficiency of \sys, we collect the time overhead of the feedback monitoring and the feedback-driven search of \sys in the eighteen searches on four datasets.
In addition, we also record the memory used in the repaired searches of \sys and the searches of the AutoKeras pipeline.
Notice that experiments for runtime overhead and memory overhead are conducted individually to avoid influencing each other.
All overhead is calculated as the average of each dataset.
The memory overhead and the time overhead distribution are shown in~\autoref{t:memoryoverhead}.
The bar chart in~\autoref{fig:timeoverhead} also intuitively shows the distribution of the time overhead in the searches of \sys on four datasets.

\begin{table}[]
	\centering
	\scriptsize
	\caption{The Overhead of Time and Memory of \sys in Searches}
	\label{t:memoryoverhead}
	\tabcolsep=3pt
	\begin{tabular}{crrrrrrr}
	\toprule
	\multirow{2}{*}{Dataset} & \multicolumn{4}{c}{Memory Overhead (GB)} & \multicolumn{3}{c}{Time Overhead of \sys (\%)} \\ \cmidrule(r){2-5} \cmidrule(r){6-8}
	& \multicolumn{1}{c}{Greedy} & \multicolumn{1}{c}{Bayesian} & \multicolumn{1}{c}{Hyperband} & \multicolumn{1}{c}{\sys} & \multicolumn{1}{c}{Monitor} & \multicolumn{1}{c}{Search} & \multicolumn{1}{c}{Training} \\ \midrule
	CIFAR-100 & 4.34 & 4.27 & 4.50 & 5.60 & 1.00 & 0.08 & 98.92 \\
	Food-101 & 7.38 & 8.90 & 8.52 & 8.51 & 0.31 & 0.02 & 99.67 \\
	Stanford Cars & 7.36 & 8.85 & 9.13 & 7.79 & 1.98 & 0.07 & 97.95 \\
	TinyImageNet & 7.35 & 9.35 & 8.25 & 7.50 & 0.36 & 0.07 & 99.57 \\ \midrule
	Avg. & 6.61 & 7.84 & 7.60 & 7.35 & 0.73 & 0.05 & 99.22\\ \bottomrule
	\end{tabular}
\end{table}

\smallskip
\noindent
{\bf Runtime Overhead Analysis:}
The search time in \sys is mainly used in three parts, namely the feedback monitoring, the feedback-driven search, and the model training and evaluation process, i.e., ``Monitor'', ``Search'', and ``Training'' in~\autoref{t:memoryoverhead}.
The first two are the time overhead from the debugging and repairing process from \sys, and the latter is the time consumption of the model training and other modules in the pipeline.
As shown in~\autoref{t:memoryoverhead} and~\autoref{fig:timeoverhead}, on the four datasets, the average time overhead of the feedback monitoring is 0.73\%, and the feedback-driven search uses 0.05\% of the total search time to analyze the feedback and select the optimal actions.
It means that over 99\% of search time spends on the model training, and \sys only takes a little time to conduct effective repair for the search bugs
It is worth noting that for the large-scale datasets, such as Food-101 and TinyImageNet, the overall time overhead of the monitor and search strategy in \sys on these two datasets is only 0.33\% and 0.43\%, which are significantly smaller than the average proportion of the overhead.
It shows that for the large-scale datasets that require heavy training processes, the time overhead of \sys will be relatively small, and the efficiency of \sys will be higher.

\smallskip
\noindent
{\bf Memory Overhead Analysis:}
The column ``Memory Overhead'' in~\autoref{t:memoryoverhead} shows the average memory overhead of \sys and other search strategies in the AutoKeras engine.
The Bayesian strategy, which has the highest cost, uses 7.84 GB of memory on average across the four datasets.
We analyze its search process and find that this high memory overhead comes from the Bayesian optimization model in this strategy, which conducts the search and generates models.
The memory overhead of the Hyperband strategy is 7.60 GB, which is lower than the overhead of the Bayesian strategy.
The Greedy strategy, as the simplest search method in the pipeline, has the lowest memory overhead of 6.61 GB.
\sys spends 7.35 GB memory on average on four datasets, which is less than the memory overhead of the Hyperband strategy.
\sys requires the gradient values, training histories, and model hyperparameters as the feedback data to analyze the current search.
The feedback data is either stored as part of the tensor data for the training purpose or automatically collected by the pipeline.
The overhead caused by \sys is mostly due to program variables, which are negligible compared with neuron and gradient values.
This makes the memory overhead of \sys only slightly larger than that of the Greedy search, which is the simplest strategy, but it still has a relatively small overhead than other search methods (e.g., Bayesian) of the AutoKeras pipeline.

\subsection{Ablation Study}\label{s:ablation}

\noindent
{\bf Experiment Design and Results:}
\sys provides three mechanisms to repair the performance bug and ineffective search bug in the AutoML engines, which are the search space expansion, the feedback-driven search, and the feedback monitoring.
To understand the contribution of the solutions in \sys, we conduct two ablation experiments on the CIFAR-100 and TinyImageNet datasets to study the performance of the search space expansion and the feedback-driven search in repairing.
Considering that the feedback-driven search fixes the bugs together with the feedback monitoring, we treat these two mechanisms as one solution in the experiments.
The first experiment uses the three search methods of the AutoKeras pipeline to search on the expanded search space to study whether the search space expansion in \sys helps find optimal models and improves the search performance.
In the second experiment, we use the original search space in all searches and only compare the feedback-driven search strategy with three AutoKeras search methods to explore whether the feedback-driven strategy can conduct more effective and efficient searches.
Both of the ablation experiments search on the dataset for at least 25 trials and 24 hours, and other configurations follow the default settings in~\autoref{s:setup}.

\begin{figure}[]
	\centering
	\footnotesize
	\begin{subfigure}[t]{0.47\columnwidth}
		\centering     
		\footnotesize
		\includegraphics[width=\columnwidth]{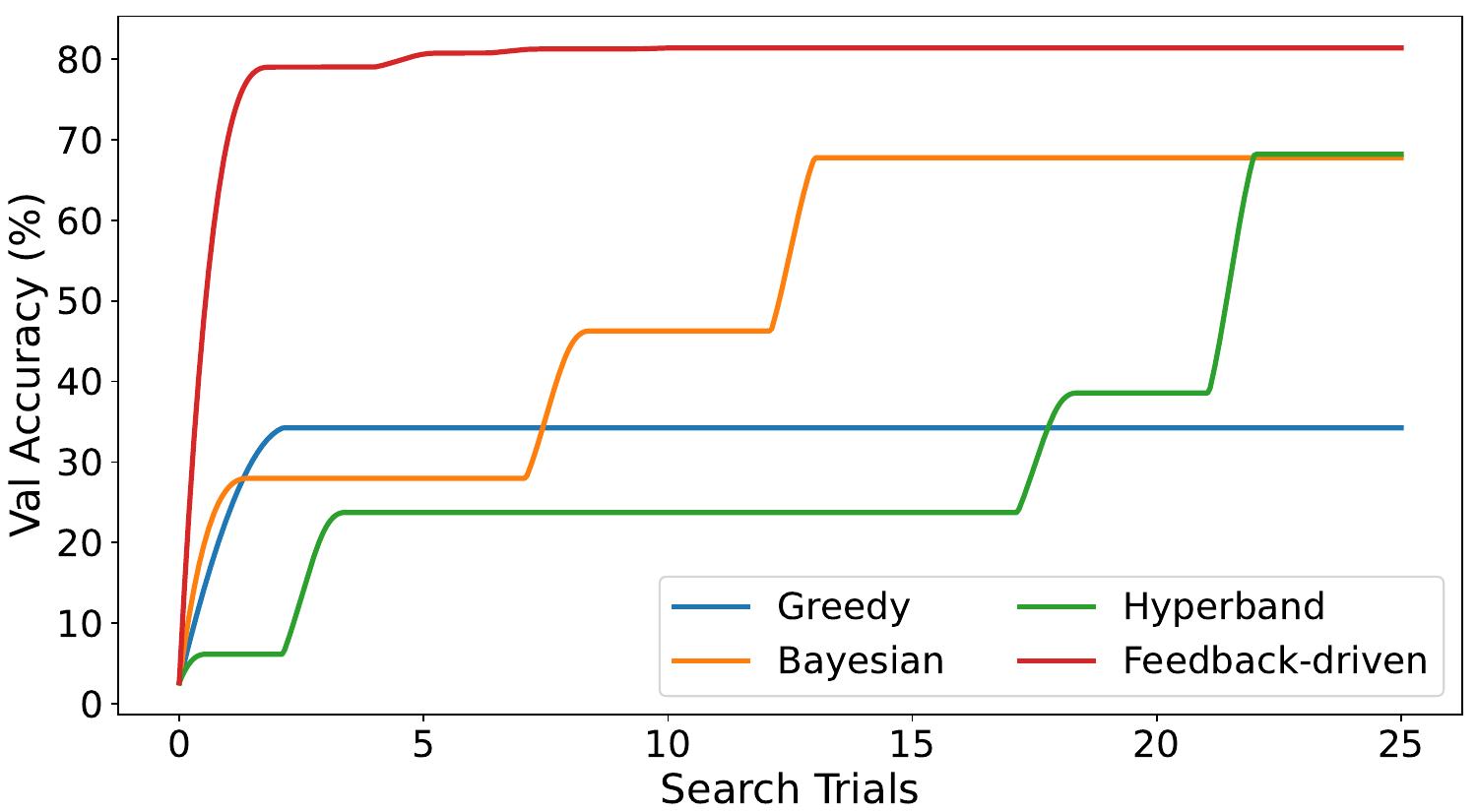}
		\caption{CIFAR-100 Dataset}
		\label{fig:rq3_cifar100}
	\end{subfigure}
	\hfill
	\begin{subfigure}[t]{0.47\columnwidth}
		\centering     
		\footnotesize
		\includegraphics[width=\columnwidth]{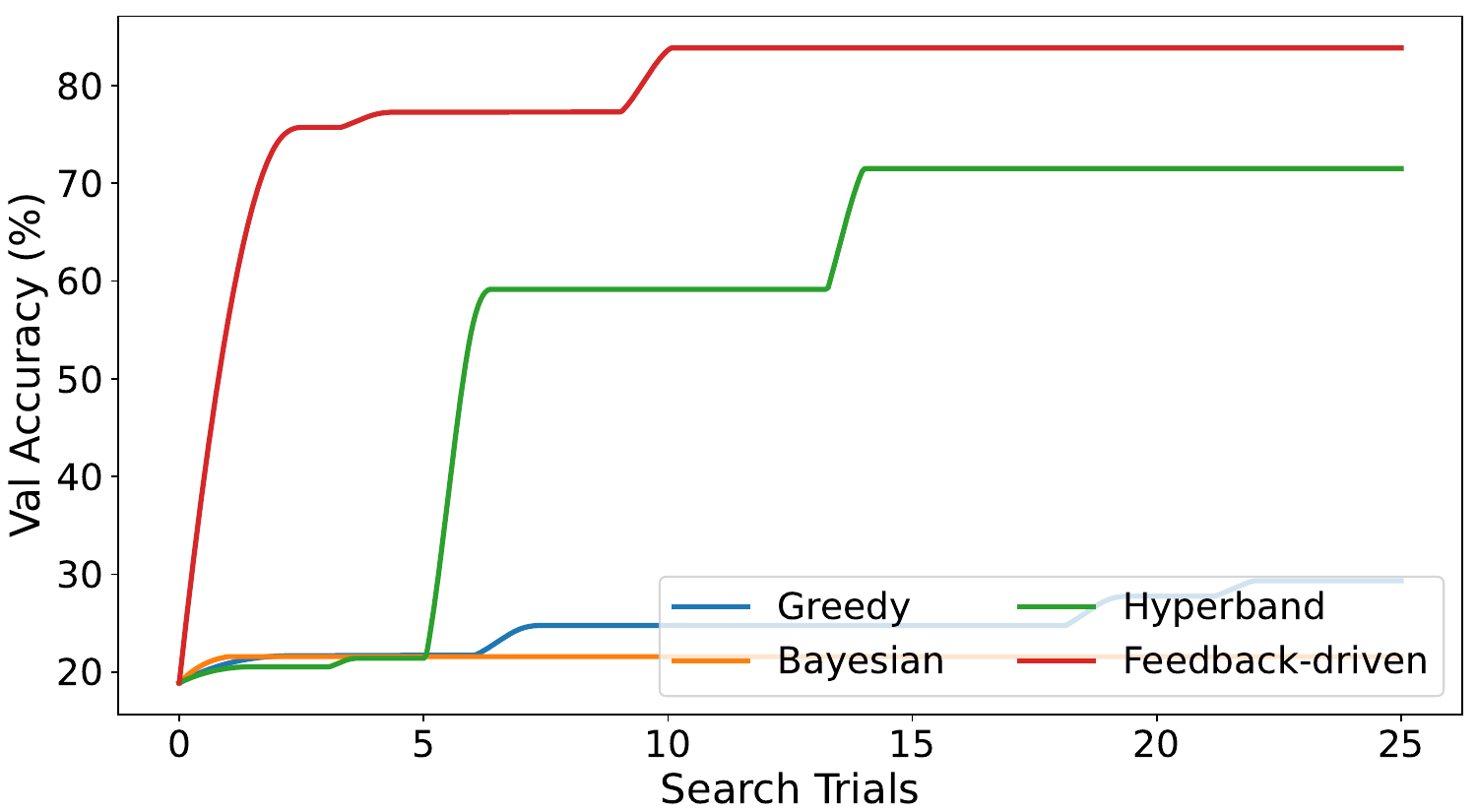}
		\caption{TinyImageNet Dataset}
		\label{fig:rq3_tiny}
	\end{subfigure}
	\caption{Ablation Study of Feedback-driven Search on two Datasets}
	\label{fig:rq3_method}
	\vspace{-10pt}
\end{figure}

\autoref{t:rq3_1} shows the ablation experiment results of the search space expansion.
The first three columns in this table show the dataset, search methods, and the score of initial models.
The column of ``Best Score'' lists the best score each search reaches in searching, and ``Score Improvement'' shows the absolute accuracy improvement that the best score achieves.
The column ``Orig.'', ``Expan.'', and ``Ratio'' separately display the search results on the original search space and the expanded search space, and the ratio between the two results.
In addition, \autoref{fig:rq3_method} shows the ablation experiment results of the feedback-driven search.
The X-axis in the figure is the number of the searched trials, and the Y-axis is the best score that each search achieves.


\begin{table}[]
	\centering
	\scriptsize
	\caption{Ablation Study Results of Search Space Expansion}
	\label{t:rq3_1}
  \tabcolsep=2.5pt
	\begin{tabular}{ccrrrrrrr}
	\toprule
	\multirow{2}{*}{Dataset} &
	\multirow{2}{*}{\begin{tabular}[c]{@{}c@{}}Search\\ Method\end{tabular}} &
	\multicolumn{1}{c}{\multirow{2}{*}{\begin{tabular}[c]{@{}c@{}}Initial\\
	Score (\%)\end{tabular}}} & \multicolumn{3}{c}{Best Score} &
	\multicolumn{3}{c}{Score Improvement} \\ \cmidrule(r){4-6} \cmidrule(r){7-9}
	&  & \multicolumn{1}{c}{} & \multicolumn{1}{c}{Orig. (\%)} &
	\multicolumn{1}{c}{Expan. (\%)} & \multicolumn{1}{c}{Ratio} &
	\multicolumn{1}{c}{Orig. (\%)} & \multicolumn{1}{c}{Expan. (\%)} &
	\multicolumn{1}{c}{Ratio} \\
	\midrule
	& Greedy & 20.73 & 34.23 & 83.66 & 2.44 & 13.50 & 62.94 & 4.66 \\
	& Bayesian & 22.26 & 67.79 & 80.47 & 1.19 & 45.53 & 58.21 & 1.28 \\
	& Hyperband & 4.68 & 68.21 & 82.70 & 1.21 & 63.53 & 78.03 & 1.23 \\
	\multirow{-4}{*}{CIFAR-100} & Avg. & 15.89 & 56.74 & 82.28 & 1.45 & 40.85 & 66.39 & 1.63 \\ \midrule
	& Greedy & 21.58 & 29.34 & 81.40 & 2.77 & 7.76 & 59.82 & 7.71 \\
	& Bayesian & 22.24 & 22.24 & 78.07 & 3.51 & 0.00 & 55.84 & - \\
	& Hyperband & 19.94 & 71.50 & 81.28 & 1.14 & 51.56 & 61.34 & 1.19 \\
	\multirow{-4}{*}{\begin{tabular}[c]{@{}c@{}}Tiny\\ ImageNet\end{tabular}} & Avg. & 21.25 & 41.02 & 80.25 & 1.96 & 19.77 & 59.00 & 2.98 \\ \midrule
	\multicolumn{2}{c}{Avg.} & 18.57 & 48.88 & 81.27 & 1.66 & 30.31 & 62.70 & 2.07 \\ \bottomrule
	\end{tabular}
\end{table}

\begin{figure*}
    \centering
  \begin{subfigure}[figure1]{0.23\textwidth}  
        \centering
    \includegraphics[width=\textwidth]{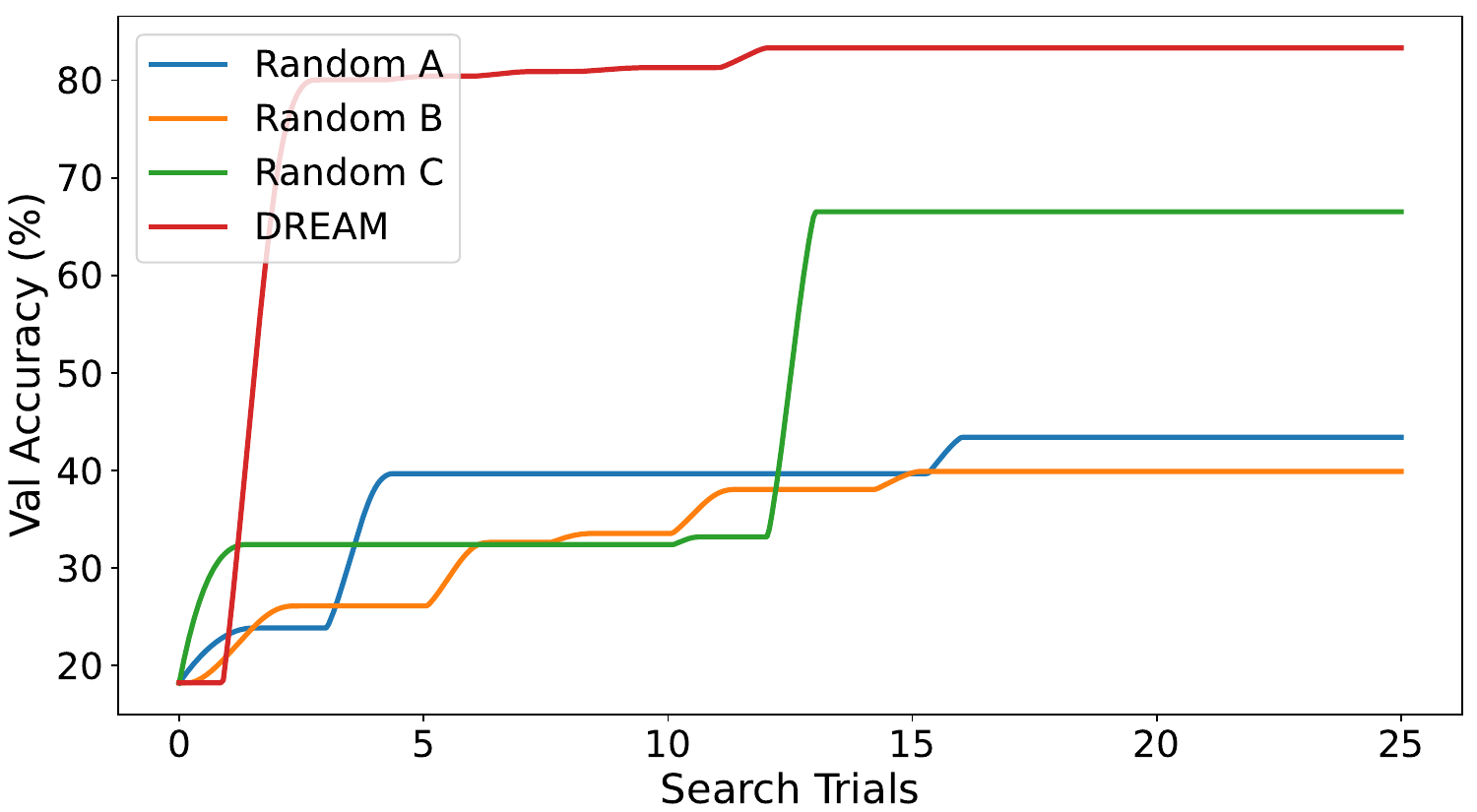} 
      \caption{CIFAR-100 Dataset}
      \label{f:rq4_cifar100}
  \end{subfigure}
  \begin{subfigure}[figure2]{0.23\textwidth}
        \centering
    \includegraphics[width=\textwidth]{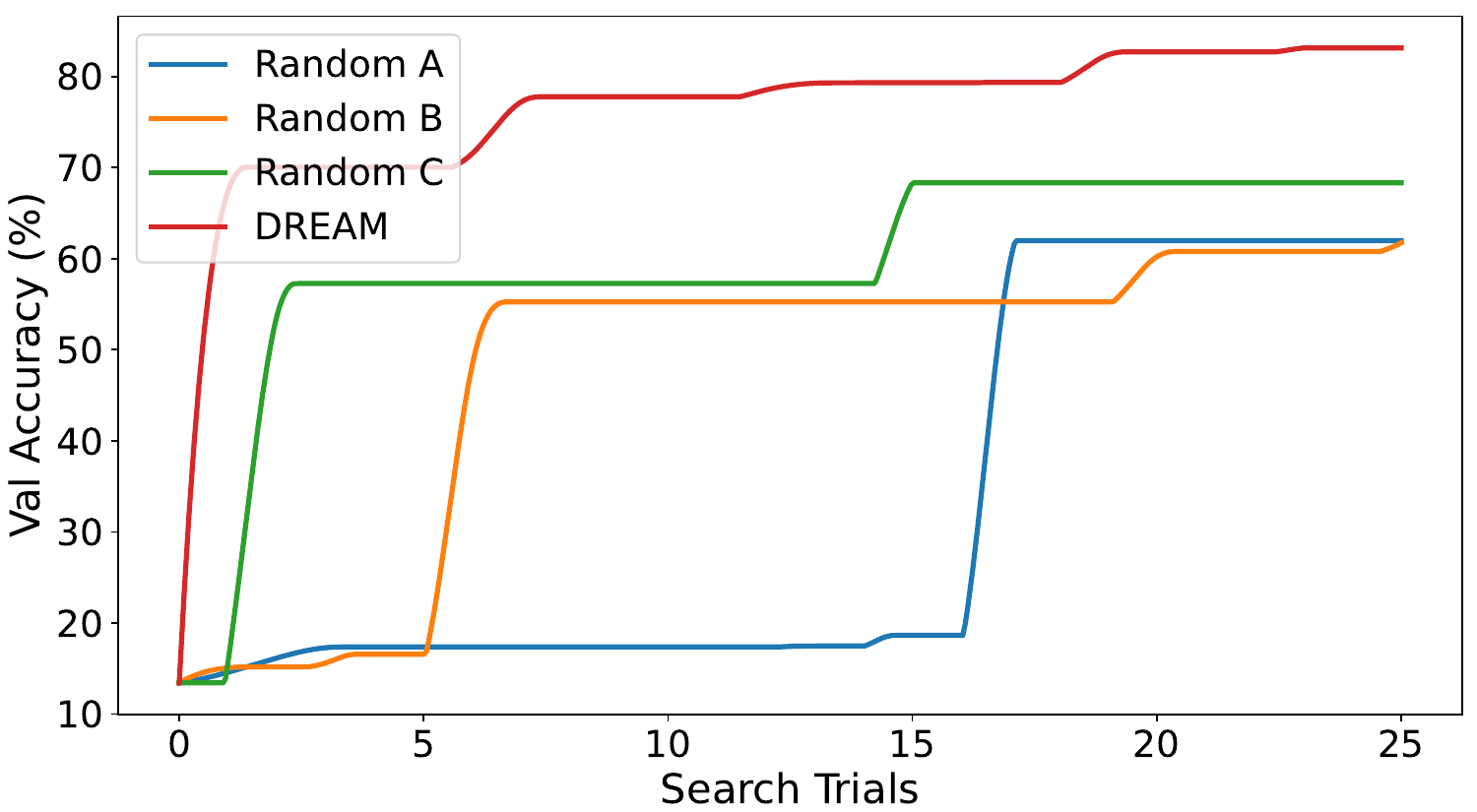} 
      \caption{Food-101 Dataset}
      \label{f:rq4_f101}
  \end{subfigure}
  \begin{subfigure}[figure3]{0.23\textwidth} 
        \centering
    \includegraphics[width=\textwidth]{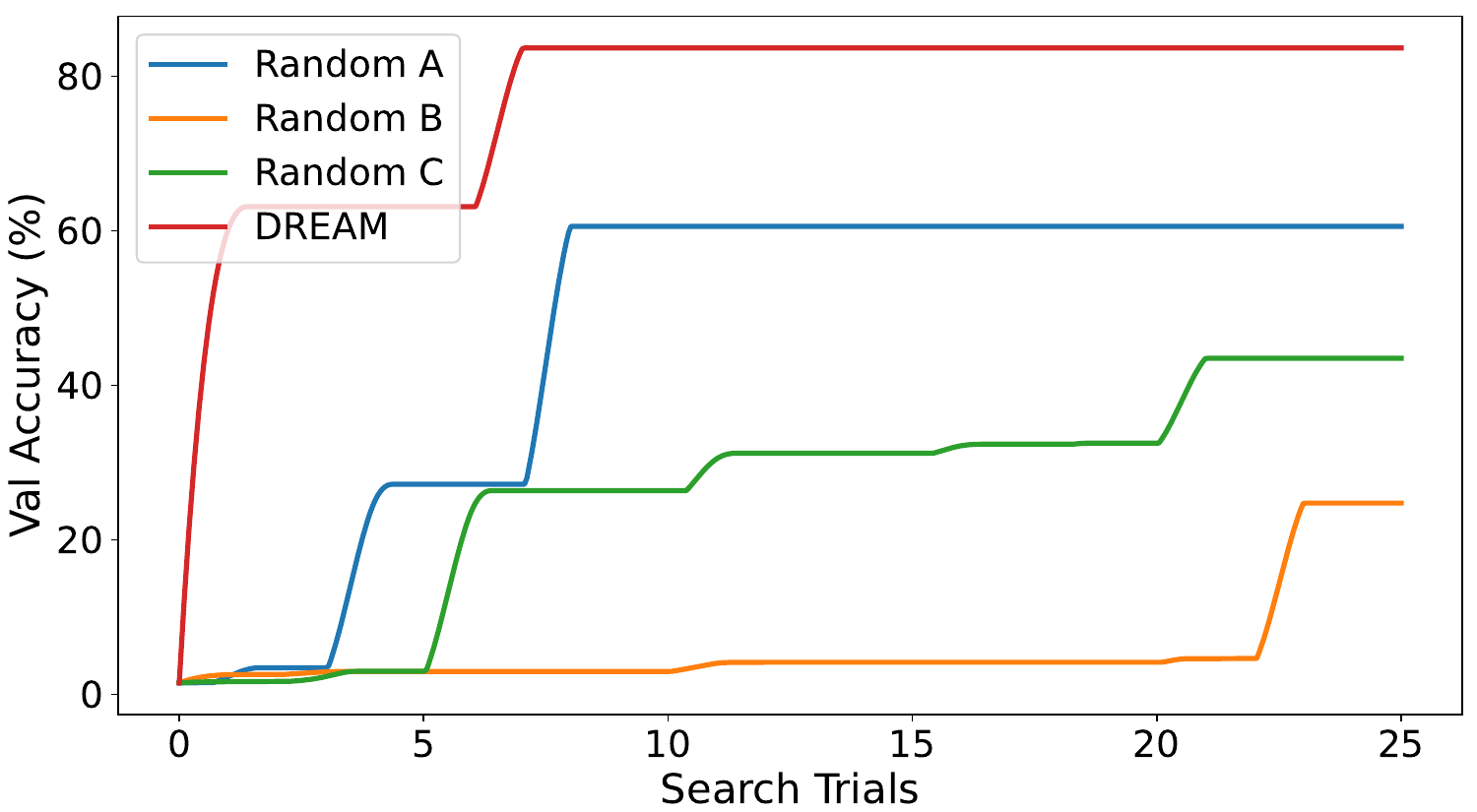} 
      \caption{Stanford Cars Dataset}
      \label{f:rq4_cars}
  \end{subfigure}
  \begin{subfigure}[figure4]{0.23\textwidth}
        \centering
    \includegraphics[width=\textwidth]{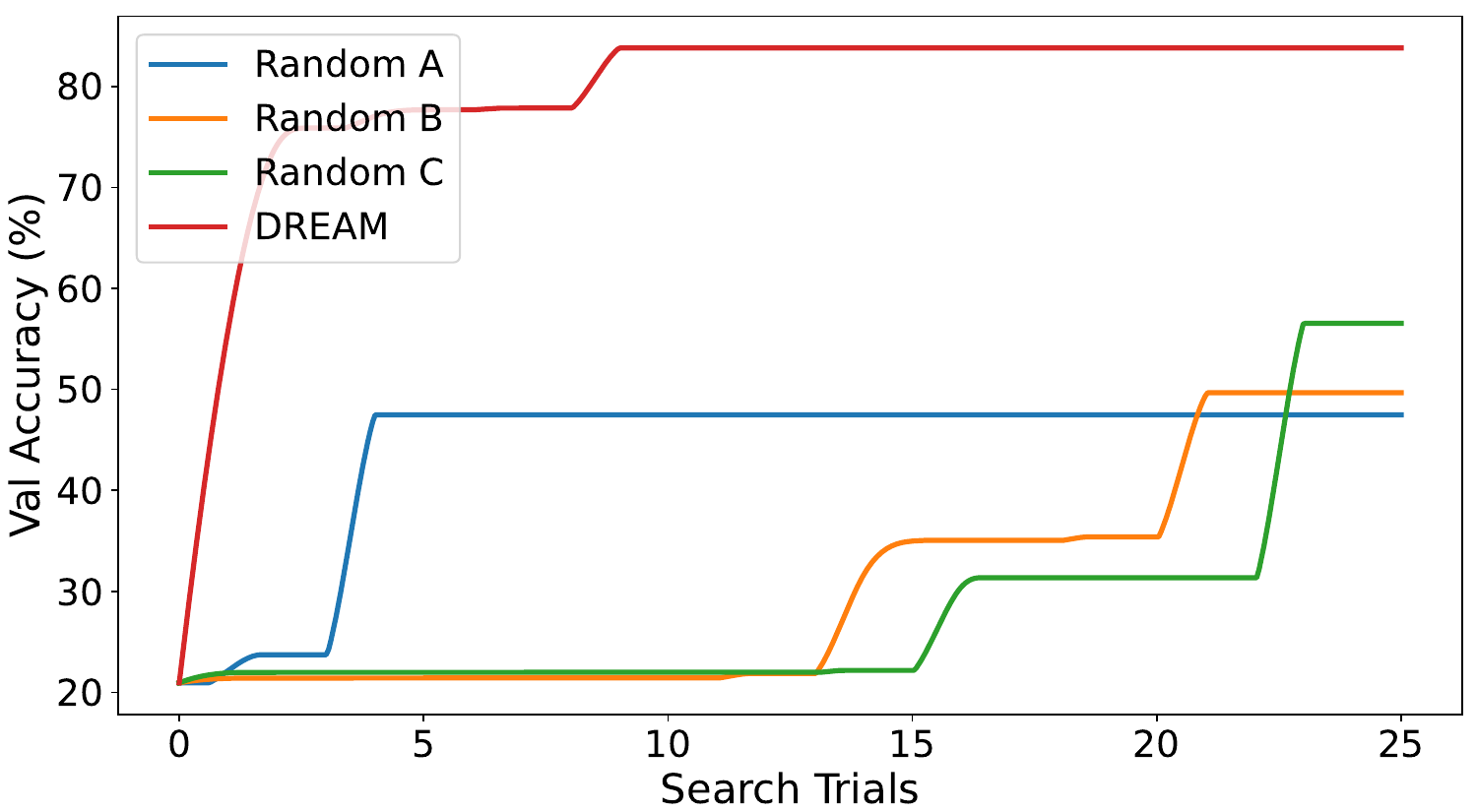} 
      \caption{Tiny ImageNet Dataset}
      \label{f:rq4_tiny}
  \end{subfigure}
  \caption{The Search Results With Different Priorities on Different
  Datasets.}
  \label{fig:rq4}
  \vspace{-10pt}
\end{figure*}

\smallskip
\noindent
{\bf Analysis:}
The experiment results illustrate the effectiveness of each solution in \sys in searching on given tasks.

\autoref{t:rq3_1} shows that the expanded actions in the search space can lead to much better models on the given datasets.
It achieves an average accuracy of 81.27\% on the six searches with three search strategies of the AutoKeras pipeline, which is significantly better than the 48.88\% accuracy of the searches on the original search space.
On the CIFAR-100 dataset, the searches on the expanded search space have achieved an average score of 82.28\%, which is 1.45 times the score of the searches on the original search space.
On the TinyImageNet dataset, the search on the expanded search space improves the initial score by 59.00\%, which is 2.98 times the improvement of the original search space.
In addition, we can observe that it is difficult for the search strategies to achieve the target accuracy rate of 70\% in the original search space, let alone 80\%.
In contrast, after implementing the search space expansion, almost all methods reach 80\% accuracy with limited search trials and search time, illustrating the role of this mechanism in fixing search performance problems.
We manually analyzed the best models found on the expanded search space.
The analysis results indicate that all these models use additional training optimization search actions, such as fine-tuning with two-step training and three-step training.
These expanded search actions greatly improve the model performance and help the models achieve accuracy far exceeding those of the models on the original search space.
The models with excellent performance demonstrate the positive impact of the search space expansion in repairing the ineffective searches bug of the pipeline and finding optimal models.

From~\autoref{fig:rq3_method}, we can observe that the feedback-driven search has significant improvement on the effectiveness and efficiency of search.
The feedback-driven search can quickly improve the accuracy to about 80\% within several trials, while the search strategies of AutoKeras are difficult to achieve 70\% accuracy during the whole 25 trials.
Take~\autoref{fig:rq3_cifar100} for example, the feedback-driven search process first searches for the actions that change the model architectures to \textit{XceptionNet} and enable the trainable setting of the pre-trained model, which improves the model accuracy to 68.77\%.
Then it changes the value of \texttt{imagenet size} parameter in the model and applies other actions according to the priority under the current conditions to further improve the search score, and finally breaks the 80\% accuracy in the fourth trial.
Notably, the feedback-driven search takes 0.61 hours to reach 70\% accuracy, while Hyperband, the best performing search strategy in AutoKeras, requires 5.45 hours.
The Bayesian search strategy doesn't even reach the target score in 43 hours of search.
In experiments, the efficiency and effectiveness of feedback-driven search considerably surpass the AutoKeras search methods, demonstrating the effectiveness of this solution in fixing the ineffective search and performance bugs.



\subsection{Effects of Different Priorities}\label{s:priority}

\noindent
{\bf Experiment Design and Results:}
\sys leverages a built-in search priority to determine the optimal action to improve the search under different conditions, which plays a vital role in the repaired search process.

To verify the effect of the search priority, we conduct a comparative experiment with different priority settings on four datasets.
In the experiment, we randomly generate three different search priorities to compare with the built-in search priority in \sys.
The built-in priority in \sys is from the conditional probability in~\autoref{sec:priority_analyzer}.
Each search contains 25 trials, and the other configurations follow the description in~\autoref{s:setup}.
The initial models and training configurations of the searches with different priorities are the same.

The search results of different priorities on four datasets are shown in~\autoref{fig:rq4}.
From these figures, we can observe that the search effect of \sys is significantly better than the effect based on the random priorities.
Additionally, \sys can often obtain outstanding search results within the first ten trials.
In contrast, the searches with random priority are always difficult to improve the training effect of the models.
They often fail to achieve 60\% accuracy in dozens of hours of searching. 
This significant difference illustrates the effectiveness of the built-in search priority of \sys in repairing searches.

\smallskip
\noindent
{\bf Analysis:}
To further analyze the impact of priority on search performance, we manually analyze the searches in~\autoref{fig:rq4} to understand how the priority in \sys guides the search process.
Taking~\autoref{f:rq4_tiny} as an example, the summarized feedback of the initial model is ``RA-NC-NG-NW''.
Based on the built-in priority in \sys, the search process in \sys searches for ``block type=xception'' and ``trainable=True'' actions firstly.
These actions change the model architecture and the \texttt{trainable} hyperparameter and significantly improve the model performance from 21.49\% to 74.19\% in the second trial, and the summarized conditions of the feedback turn to ``XA-NC-NG-NW''.
Then \sys tries to change the value of \texttt{imagenet size} hyperparameter and applies the actions that change the AdamW optimizer configuration, increasing the search score to 77.67\%.
At this time, the searches with random priorities attempt to search for actions such as adjusting the learning rate to \(0.01\), but only slightly improving the scores, no more than 50\%.
Subsequently, the feedback-driven search in \sys continues to search for actions such as changing the model architecture to \textit{EfficientNet} and finally reaches 83.60\% accuracy in the 18th trial.
In contrast, the random priority-driven search with the best performance (i.e., ``Random C'' in~\autoref{f:rq4_tiny}) only obtains the best score of 56.54\% by adjusting the learning rate of Adam optimizer to \(0.001\).


\section{Conclusion}\label{sec:conclusion}

This paper presents \sys, an automatic debugging and repairing system for AutoML systems with a focus on performance and ineffective search bugs.
\sys collects the detailed feedback from the model training and evaluation and detects the AutoML bugs with the feedback monitor.
Then it repairs the bugs by expanding search space and conducting a feedback-driven search.
Our evaluation results show that \sys can effectively and efficiently fix AutoML bugs, and the repaired search obtains better results than state-of-the-art AutoML frameworks.
\newpage

\bibliographystyle{ACM-Reference-Format}
\bibliography{ref}


\end{document}